\documentclass{aa}

\usepackage{graphicx}
\usepackage{natbib}
\bibpunct{(}{)}{;}{a}{}{,}
 
\begin{document}
 
\title{On the internal structure of starless cores. I.
Physical conditions and the distribution of CO, CS, N$_2$H$^+$, 
and NH$_3$ in L1498 and L1517B}

\author{M. Tafalla \inst{1}
\and
P.C. Myers \inst{2}
\and P.Caselli \inst{3}
\and C.M. Walmsley \inst{3}
}
 
\institute{Observatorio Astron\'omico Nacional, Alfonso XII 3, E-28014 Madrid,
Spain
\and
Harvard-Smithsonian Center for Astrophysics, 60 Garden St., Cambridge,
MA 02138, USA
\and
Osservatorio Astrofisico di Arcetri, Largo E. Fermi 5, I-50125
Firenze, Italy
}
 
\offprints{M. Tafalla \email{m.tafalla@oan.es}}
\date{Received -- / Accepted -- }
 
\abstract{
We have characterized the physical structure and chemical composition of two 
close-to-round starless cores in Taurus-Auriga,
L1498 and L1517B. Our analysis is based 
on high angular resolution observations in 
at least two transitions of NH$_3$, N$_2$H$^+$, CS, 
C$^{34}$S, C$^{18}$O, and C$^{17}$O, together with maps of the 1.2 mm
continuum. For both cores, we derive radial profiles of constant 
temperature and constant turbulence, together 
with density distributions close to those
of non-singular isothermal spheres. Using these physical 
conditions and a Monte Carlo radiative transfer
model, we derive abundance profiles for all species
and model the strong chemical differentiation
of the core interiors. According to our
models, the NH$_3$ abundance increases toward the core centers
by a factor of several $(\approx 5)$ while N$_2$H$^+$
has a constant abundance over most of the cores. In contrast, both
C$^{18}$O and CS (and isotopomers) are strongly depleted in the core
interiors, most likely due to their freeze out 
onto grains at densities of a few 10$^4$ cm$^{-3}$. Concerning the 
kinematics of the dense gas, we find (in addition to constant turbulence)
a pattern of internal motions at the level of 0.1 km s$^{-1}$. 
These motions seem 
correlated with asymmetries in the pattern of molecular depletion,
and we interpret them as residuals of core contraction. 
Their distribution and size suggest that 
core formation occurs in a rather irregular manner
and with a time scale of a Myr. A comparison of our derived core 
properties with those predicted by
supersonic turbulence models of core formation 
shows that our Taurus cores are much more quiescent than 
representative predictions from 
these models.  In two appendices at the end of the paper
we present a simple and accurate approximation to the density profile
of an isothermal (Bonnor-Ebert) sphere, and a Monte Carlo-calibrated 
method to derive gas kinetic temperatures using NH$_3$ data.
}

\authorrunning{Tafalla et al.}
\titlerunning{On the internal structure of starless dense cores. I}
 
\maketitle

\section{Introduction}

Dense cores like those in Taurus and other nearby dark clouds
represent the simplest environments where stars form. Their study holds
the
promise of revealing the basic physics of star formation, but despite
their apparent simplicity, dense cores 
remain mysterious in their internal structure, 
formation and evolution, and in the manner in which they collapse under gravity 
(see, e.g., \citealt{mye99,eva99} for recent reviews). Part of this
mystery results from a number of contradictions between 
observations of different dense gas tracers (see, e.g., \citealt{zho89}),
which have so far precluded a self consistent description of the core internal 
properties. Fortunately,
multi-molecular and continuum observations over the last few years 
have started to suggest that most contradictions arise from
cores not being chemically homogeneous, as initially assumed, 
but having strongly differentiated interiors 
(\citealt{kui96,wil98,kra99,alv99,cas99,ber01,jes01,taf02,ber02,bac02}). 
Understanding and characterizing the chemical structure 
of dense cores has therefore become a top priority for star formation 
studies, as solving this problem may finally produce 
a coherent picture of dense cores, from which 
their star forming evolution may be inferred.

	If cores have strong chemical composition gradients,
in addition to density gradients, their description cannot
rely on average parameters, as often assumed.
It is necessary to use spatially-variable abundances, and these
need to be derived from the data 
through a realistic modeling of the radiative transfer.
This approach, unfortunately, involves a number of uncertainties, one of
the largest being 
our ignorance of the exact (3D) geometry of a core, which
is not well constrained either by theory or 2D images (e.g., 
compare \citealt{jon02} with \citealt{cur02}). To minimize
this effect, it is convenient to concentrate on cores with relatively
simple geometries, and for this reason, we have selected L1498 and L1517B
as subjects of the present study. 

	L1498 and L1517B are two dense cores in the Taurus-Auriga
cloud complex, and were first identified by \citet{my83b} from the
inspection of Palomar Sky Atlas prints. They are not associated with
IRAS point sources \citep{bei86} or 1.2 millimeter point
sources \citep{taf02}, and can therefore be considered starless.
Over the last decade, L1498 has been the subject of a number of
detailed studies: \citet{lem95} mapped its C$^{18}$O and CS emission, 
\citet{kui96} first identified it as a chemically differentiated system, 
and further details of its chemical structure have been presented by 
\citet{wol97} for CCS and \citet{wil98} for CO. More recently, 
\citet{lan01} have presented an estimate of the core density profile from
FIR data and \citet{lev01} have estimated an upper limit to
its magnetic field of 100~$\mu$G from Zeeman observations. 
Tafalla et al. (2002, TMCWC hereafter) presented line and
continuum observations of L1498, 
L1517B, and three other dense cores, and found a systematic pattern 
of chemical differentiation: species like CS and CO 
are almost absent from
the core interiors while NH$_3$ and N$_2$H$^+$ are present even in 
the densest parts. In this paper, we improve the study of 
L1498 and L1517B using new multiline, high angular resolution observations
and creating for each core a model of the internal conditions 
that simultaneously explains all our C$^{18}$O, 
C$^{17}$O, CS, C$^{34}$S, N$_2$H$^+$, and NH$_3$ data.
As a result, we confirm the presence of a systematic 
abundance pattern, and show that the two cores are very close to
isothermal spheres both in their density profiles and 
physical parameters (temperature and turbulence). At the same time, we 
infer that the core contraction history has been asymmetrical, 
both from the internal kinematics and the chemical composition.
In a follow-up
paper (Tafalla et al. 2004, in preparation) we will present a full 
chemical survey of these two cores.

\section{Observations}

We observed L1498 and L1517B in the 1.2 mm continuum with the IRAM 30m 
telescope in 1999 December and 2000 December. We used the MAMBO 1 
bolometer array \citep{kre98} in on-the-fly mode with  a scanning speed of
$4''$ s$^{-1}$, a wobbler period of 0.5 s, and a wobbler throw of 53$''$. 
The data were corrected for atmospheric 
extinction using sky dips done before and after
each individual map, and they were reduced using the NIC software \citep{bro96}.
A global calibration factor of 15000 counts per Jy beam$^{-1}$ 
was estimated for both 1999 and 2000 epochs from observations of Uranus. 
The bolometer central frequency is 240 GHz and the
telescope beam size is approximately $11''$. (The 1999 data have already
been presented in \citealt{taf02}.)

We observed L1498 and L1517B in C$^{18}$O(1--0), C$^{18}$O(2--1),
C$^{17}$O(2--1), CS(2--1), CS(3--2), C$^{34}$S(2--1), C$^{34}$S(3--2),
and N$_2$H$^+$(1--0) with the IRAM 30m telescope in 2000 October. L1498 was 
mapped in all lines and L1517B was mapped in all but C$^{34}$S(2--1) 
and C$^{34}$S(3--2), which were so weak that only the central position
was observed. Additional N$_2$H$^+$(3--2) observations were carried out
in 2001 April. All data were taken in frequency switching mode
with frequency throws of approximately 7 or 14 MHz. As a backend we 
used the facility autocorrelator, split into several windows to achieve
velocity resolutions between 0.03 and 0.06 km s$^{-1}$. The 
telescope pointing was corrected using frequent cross scans of continuum 
sources, and the total errors were found to be smaller than $4''$ (rms). 
The data were calibrated using the standard chopper wheel method
and converted to the main beam brightness temperature scale. The full
width at half maximum (FWHM) of the telescope beam is inversely
proportional to the frequency, and ranges from approximately
$26''$ at the (lowest) frequency of N$_2$H$^+$(1--0) to $11''$ at the (highest)
frequency of N$_2$H$^+$(3--2).

As stressed in previous studies (\citealt{lee01,taf02}), the modeling 
and interpretation of
the very narrow spectral lines from starless cores like L1498 and L1517B 
(FWHM $\approx$ 0.2 km s$^{-1}$ $\approx$  70 kHz at 100 GHz) requires using 
rest frequency values more accurate
than those in standard compilations. Fortunately, current laboratory
work is starting to produce frequency determinations with uncertainties
of the order of 1 kHz, and whenever such a measurement exists we have
preferred it over any other value, including our own previous astronomical
determinations (\citealt{lee01}). The rest frequencies used in this paper are 
indicated in Table 1.

\begin{table}
\caption[]{Rest Frequencies.
\label{tbl-1}}
\[
\begin{array}{lrcc}
\hline
\noalign{\smallskip}
\mbox{Line} & \mbox{Frequency}  & \mbox{Uncert.} & \mbox{Ref.} \\
& \mbox{(MHz)} & \mbox{(kHz)} &  \\
\noalign{\smallskip}
\hline
\noalign{\smallskip}
\mbox{C$^{18}$O(J=1--0)~~~} & 109782.172 & 3 & (1) \\
\mbox{C$^{18}$O(J=2--1)} & 219560.354 & 1 & (2) \\
\mbox{C$^{17}$O(J,F=1,7/2--0,5/2)~~~} & 112358.990 & 1 & (3) \\
\mbox{C$^{17}$O(J,F=2,9/2--1,7/2)} & 224714.214  & 1 & (3) \\
\mbox{CS(J=2--1)}        & 97980.953 & 1 & (4) \\
\mbox{CS(J=3--2)}        & 146969.026 & 1 & (4) \\
\mbox{C$^{34}$S(J=2--1)} & 96412.952 & 1 & (4)  \\
\mbox{C$^{34}$S(J=3--2)} & 144617.101& 1 & (4) \\
\mbox{NH$_3$(JKF$_1$F=1110.5--1111.5)} & 23694.501 & 0.1 & (5) \\
\mbox{N$_2$H$^+$(JF$_1$F=101--021)} & 93176.258 & 10 & (6) \\
\mbox{N$_2$H$^+$(JF$_1$F=344--233)} & 279511.811 & 6 & (7) \\
\noalign{\smallskip}
\hline
\end{array}
\]
\begin{list}{}{}
     \item[References:] (1) C. Gottlieb, priv. comm.; (2) \citet{mul01};
(3) \citet{caz02}; (4) \citet{got03}; (5) \citet{kuk67}; 
(6) \citet{lee01}, astronomical; (7) \citet{cas02c}, astronomical
\end{list}
\end{table}

\section{Results}

\subsection{Continuum data and density profiles}

\begin{figure*}
\centering
\resizebox{15cm}{!}{\includegraphics{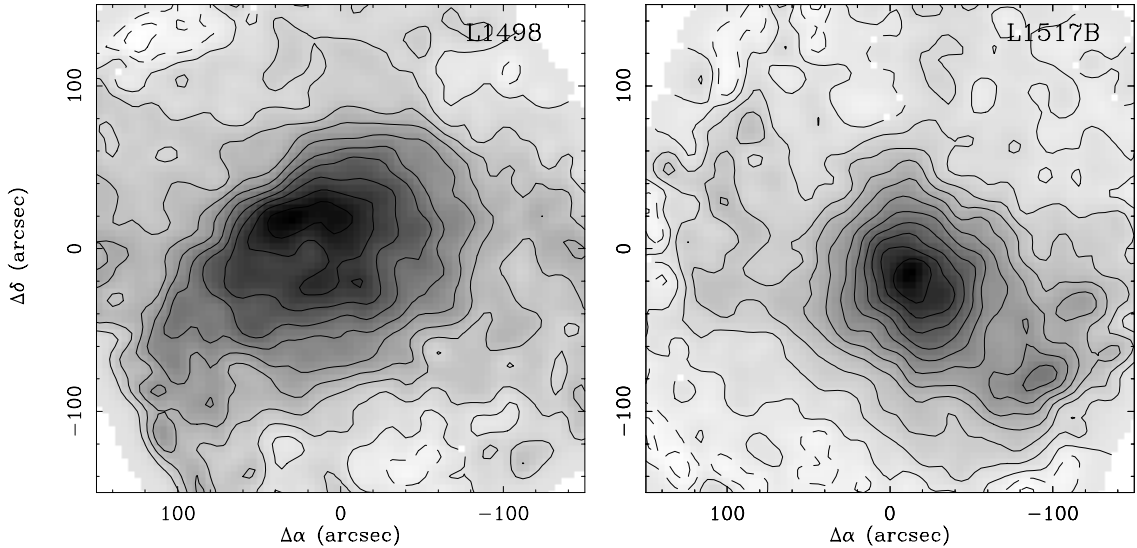}}
\caption{1.2 mm continuum maps of L1498 (left) and L1517B (right).
Note the central concentration and symmetry of the emission. In order to 
enhance the sensitivity, each map has been convolved with 
a $20''$ gaussian. First contour and contour spacing are 2 mJy/$11''$-beam,
and central coordinates are 
($\alpha_{1950}=4^{\rm h}7^{\rm m}50\fs0$, 
$\delta_{1950}=25\degr02'13\farcs0$) for L1498 and 
($\alpha_{1950}=4^{\rm h}52^{\rm m}7\fs2$, 
$\delta_{1950}=30\degr33'18\farcs0$) for L1517B.
\label{fig1}}
\end{figure*}

In contrast with the molecules, whose emissivity is highly
sensitive to excitation and chemistry (see below), the dust
in a starless core is expected to have an emissivity approximately constant 
with radius (see also below for caveats). This property allows 
a rather straightforward conversion
of the observed continuum emission into a density profile
(e.g., \citealt{war94,and96,eva01}), and for 
this reason, we start our analysis of L1498 and L1517B by
studying their dust emission.

Figure 1 presents 1.2 mm continuum maps of L1498 and L1517B, and shows 
that in both cores the mm emission 
is centrally concentrated. In L1498, the emission appears slightly 
elongated SE-NW, while L1517B is rounder with a weak extension to the SW
(and a weaker one to the NE). Neither of the cores presents evidence for 
an unresolved point-like component, indicative of an embedded object, so
they appear as bona fide starless cores.

To estimate the density distributions of L1498 and L1517B from the
data shown in Fig. 1, we follow the same procedure as in 
TMCWC (where the December 1999 part of this dataset was presented).
Thus, we approximate each core as a spherical system, and derive
a continuum radial profile by taking an azimuthal average of
its map. As L1498
is slightly elongated, we average the data of this source
in ellipses of position
angle $-40^\circ$ and aspect ratio $b/a=0.6$, while for the rounder L1517B,
we average the data in circles. These radial profiles are
presented in Fig. 2 both in linear and log-log scales.
(Core centers are assumed to be at ($10''$, 0) and ($-15''$, $-15''$)
for L1498 and L1517B, respectively.)

To model each continuum radial profile we need to assume 
a dust temperature law. Theoretical models of 
the dust temperature in cores
predict a slight inward decrease 
\citep{mat83,eva01,zuc01}, which for the case of
the L1498 and L1517B cores is estimated to range between
11 K in the outer layers and 8 K in the innermost part
(\citealt{eva01}, their model for L1512). Here, for
simplicity (and from our NH$_3$ analysis in Section 3.3), 
we will assume that both cores have a constant dust
temperature of 10 K, and we will use this value to estimate their
density radial profiles (see also \citealt{lan01}).

In addition, we assume a constant dust emissivity of
$\kappa_{1.2 mm}= 0.005$ cm$^2$ g$^{-1}$, for consistency with previous
work (e.g., \citealt{and96}). Following TMCWC, we fit the 
data using a density profile of the form
$$n(r) = {n_0\over 1+(r/r_0)^\alpha}, $$
where $n_0$ is the central density, $r_0$ is the radius of the 
inner ``flat'' region
($2 r_0$ is the full width at half maximum), and $\alpha$ is
the asymptotic power index. For each choice of $n_0$, $r_0$, 
and $\alpha$, we calculate the emergent
continuum distribution by integrating the equation of radiative
transfer in the optically thin case, and simulate an 
on-the-fly bolometer observation of such a distribution
using the same instrumental parameters as those used at the telescope.
The radial profile of this simulated observation is compared
with the measured radial profile, and the free parameters  $n_0$, $r_0$,
and $\alpha$ are modified until model and data agree.
As shown by the solid lines in Fig. 2, the final choice of free parameters
(Table 2)
gives very reasonable fits for both cores, and indicates that L1517B
is twice as dense at its center than L1498, while it has a FWHM 
approximately half that of L1498 (so both cores have similar
central column densities: 3-4$ \times 10^{22}$ cm$^{-2}$).
We note that the fit for L1517B is the same as derived in TMCWC,
while the fit for L1498 is only slightly different (by less than
10\% for radii smaller than $110''$ and by less than 30\% 
in the range 110-150$''$). 

\begin{figure}
\centering
\resizebox{\hsize}{!}{\includegraphics{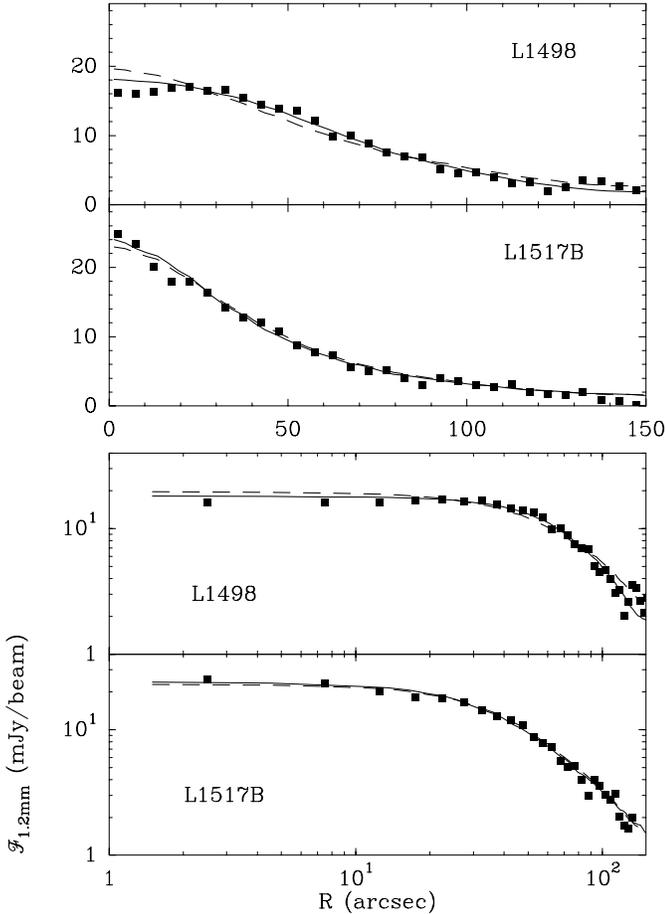}}
\caption{Radial profiles of 1.2 mm continuum emission (from the 
maps in Fig. 1). 
The squares represent observed data and the lines are the
predictions from two density models: the solid line is a simple
analytic model and the dashed line is an isothermal model (see text).
Both linear-linear and log-log plots are shown.
\label{fig2}}
\end{figure}

Although our analytical density distributions 
accurately fit the continuum data, their motivation is purely
phenomenological. As detailed in Appendix A, however,
a particular choice of parameters in this density family 
provides an excellent approximation
to an isothermal function (or Bonnor-Ebert sphere), and this choice
happens to be the best fit for L1517B (see Appendix A for further details). 
This ``coincidence'' motivated us to explore isothermal
fits to the L1517B and L1498 radial profiles
as an alternative to the analytical models presented before. For this, 
we have used the numerical solution to the isothermal function 
by \citet{cha49}, and searched for the best fit to each core 
by varying the central
density and the isothermal sound speed as independent free parameters.
In this procedure, we have followed the steps explained in the fitting with 
analytical models including the on-the-fly simulation, and
the best fit parameters are given in Table 2. As can be seen in 
Fig. 2, the isothermal models (dashed lines) provide reasonable
fits to the observed continuum data, especially for the rounder L1517B, 
where the isothermal and analytic models are
indistinguishable (see also Appendix A). For the more elongated L1498 core, 
the isothermal model does not
fit the radial profile as well as the analytic function, 
but its deviation from the data is never very large ($\sim$ 25\% at the 
center), and is less than the deviation of this core from 
spherical geometry (about 40\%). Our fits of L1498 and L1517B, therefore,
add to recent evidence that the density profiles of 
starless cores can be modeled exactly or approximately
by isothermal spheres \citep{bac00,alv01,eva01}. 

As shown by \cite{ebe55} and \cite{bon56}, a self gravitating isothermal 
sphere is stable only if it is bounded from the outside 
and its center-to-edge density contrast does not exceed a
limiting value close to 14, or equivalently, its dimensionless radius 
($\xi$) does not
exceed the critical value 6.45. Although neither L1498 
nor L1517B show a clear boundary at which the core merges with
an external medium, we can assess possible core stability
by calculating the critical radius for each core.
For the less centrally concentrated L1498 core, the
critical radius occurs at $150''$, which is at the very edge of our map
(defined by the sensitivity of our continuum measurements).
For L1517B, the critical radius occurs at $100''$; so if the isothermal
core continues to the edge of our map, the core would seem unstable.
This situation is similar to that found by \citet{bac00} in their
study of a sample of starless cores with ISOCAM. We defer further discussion 
of the isothermal model and its stability to section 4.1.

\begin{table}
\caption[]{Density fits from continuum data
\label{tbl-2}}
\[
\begin{array}{lcccccc}
\hline
\mbox{} & \multicolumn{3}{c}{\mbox{Analytic}^{\mathrm{(a)}}} & 
\mbox{~~~~} & \multicolumn{2}{c}{\mbox{Isothermal}} \\
\mbox{} & n_0 \mbox{~(cm$^{-3}$)} & r_0 \mbox{~}('')  & \alpha
& \mbox{} & n_0 \mbox{~(cm$^{-3})$} & a \mbox{~(km s$^{-1}$)}  \\
\noalign{\smallskip}
\hline
\noalign{\smallskip}
\mbox{L1498~~~~~} & 0.94 \; 10^5 & 75 & 3.5  & & 1.35 \; 10^5 & 0.32 \\
\mbox{L1517B} & 2.2 \; 10^5 & 35 & 2.5  & & 2.2 \; 10^5 & 0.27 \\
\noalign{\smallskip}
\hline
\end{array}
\]
\begin{list}{}{}
 \item[$^{\mathrm{a}}$] $n(r) = {n_0/(1+(r/r_0)^\alpha)} $  
\end{list}
\end{table}

\subsection{Overview of the molecular line data and model parameters}

\begin{figure*}
\centering
\resizebox{10cm}{!}{\includegraphics{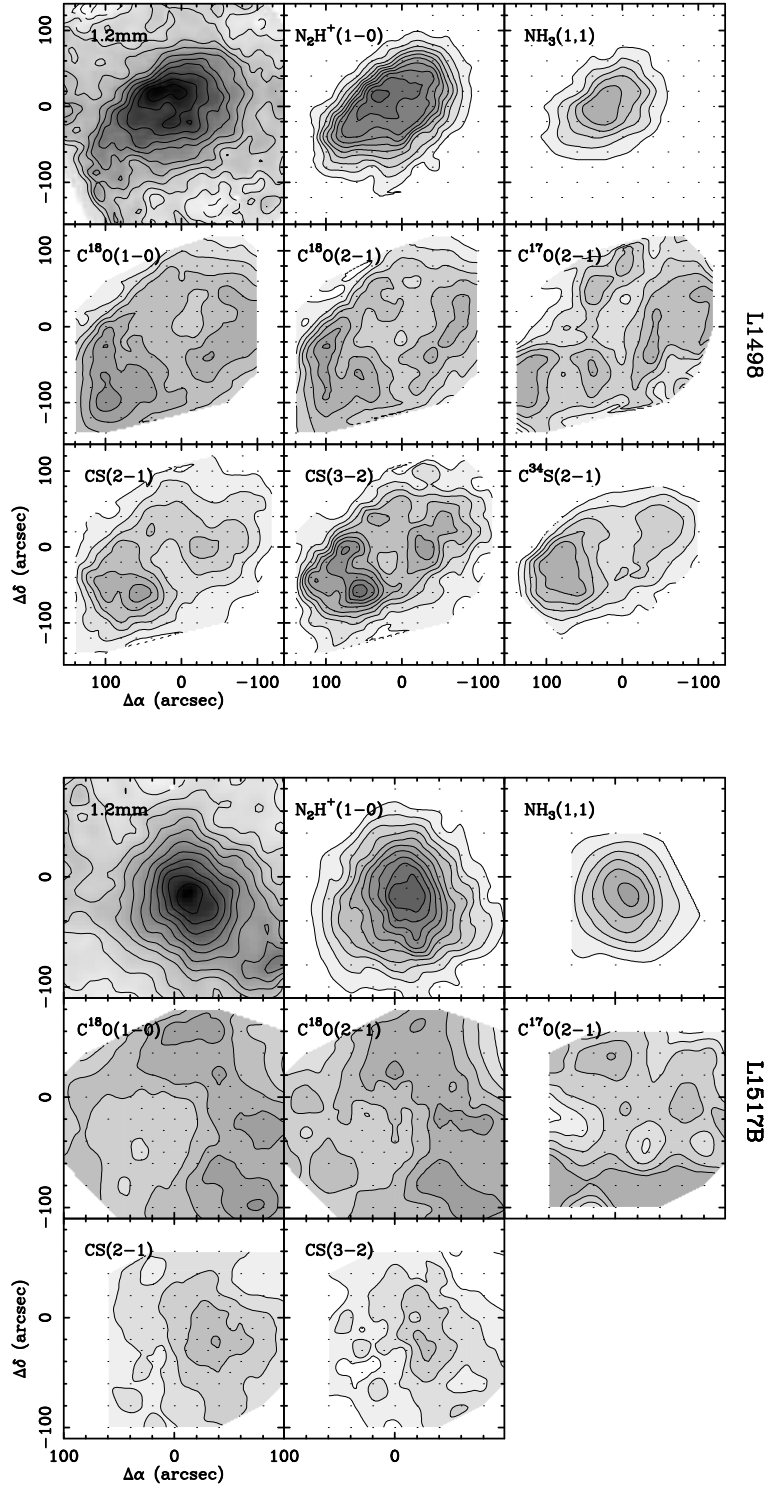}}
\caption{Integrated intensity maps for continuum and lines
observed toward L1498 and L1517B. Note the systematic
dichotomy between centrally-peaked and centrally-depressed
morphologies. For each core, the top row shows the maps of 
centrally-peaked tracers, 1.2 mm continuum, N$_2$H$^+$, and NH$_3$, and
the two lower rows present the maps of centrally-depressed species,
isotopomers of CO (middle row) and isotopomers of CS (bottom row).
In each map, the lowest contour and the contour interval are equal, and 
for each line, the same contour choice has been used for both L1498
and L1517B. Lowest contours are (in K km s$^{-1}$): 0.15 for C$^{18}$O(1--0)
and (2--1), 0.05 for C$^{17}$O(2--1), 0.25 for N$_2$H$^+$(1--0), 1.5 for
NH$_3$(1,1), 0.15 for CS(2--1), 0.1 for CS(3--2), and 0.05 for C$^{34}$S(2--1).
Continuum maps and central positions as in Fig. 1.
The C$^{17}$O(2--1) data have been convolved with a $35''$ (FWHM) gaussian
to improve S/N.
\label{fig3}}
\end{figure*}

To derive the temperature, kinematics, and chemical structure of L1498
and L1517B, we now analyze their molecular emission. Before
starting the detailed radiative transfer modeling, we first present a 
comparison between the integrated maps of all species and identify
the most general trends. As Figure 3 shows (see also TMCWC), the maps
of each core 
display a striking dichotomy ---almost to the point of anti correlation---
between the distribution of the different tracers: 
the 1.2 mm continuum, N$_2$H$^+$, and
NH$_3$ emission on the one hand
are compact and centrally concentrated, while the CO and CS isotopomer
emission on the other hand
are extended and almost ring-like. In the larger L1498 core, the
CO/CS ring is well resolved spatially, and it appears fragmented with 
a bright peak toward the SE. In the more compact L1517B core,
the CO isotopomers form a well-defined half ring toward the west, while
the CS emission is more concentrated, although it is still offset toward 
the west from the 1.2 mm/N$_2$H$^+$ maximum. 

As modeled in detail in the next sections, 
the ring distributions of the CO and CS isotopomers result from a
sharp drop in the abundances of these molecules toward the dense centers
of the cores, most likely caused by the freeze out of 
these molecules onto dust grains (for additional cases, see
\citealt{kui96,wil98,kra99,alv99,cas99,ber01,jes01,bac02}; TMCWC).
N$_2$H$^+$ and  NH$_3$, on the other hand, seem immune to this process
at typical core densities
and they are 
present in the gas phase all throughout the core. 

The goal of the following sections is to derive self consistent models
of the internal structure of 
L1498 and L1517B that fit simultaneously all line observations.
For this, we model the cores as spherically symmetric systems and 
predict  the emission from each molecule with a Monte Carlo
radiative transfer code based on that of \citet{ber79}. The physical 
properties of these models are fully defined by radial profiles of density,
temperature, turbulent velocity, and systematic velocity, and the same
parameters are used to model all molecular lines. As core density
profiles, we use the analytic functions derived from the continuum data,
and the rest of the gas parameters are derived from the molecular line data.
In our derivation of these parameters we have favored simplicity over 
detailed fine tuning, and here we summarize the main results (see following
sections for more details and Table 3 for a parameter list). To derive 
gas temperature profiles, we have used the NH$_3$ data, and we have found
that this parameter is
constant with radius in both cores (section 3.3). The turbulent profiles
are also derived from the NH$_3$ data, and they are constant with
radius, as indicated by the radial profiles of linewidth (section 3.7.1). 
The profiles of systematic velocity have been derived
from the combined fit to all line shapes, as different species
trace the gas velocity at different radii depending on optical
depth. For both L1498 and L1517B, we find that the velocity field in 
the core central part 
is constant, as indicated by the NH$_3$ data, and that there is a 
slight gradient toward the outside traced by the velocity of
the CS self absorptions
(section 3.6). For simplicity we have modeled this outer gradient with a
linear function.

When modeling the self absorbed lines of CS (section 3.6), HCO$^+$, and
HCN (Tafalla et al. 2004 in preparation), the shape of the profiles
depends very sensitively on the gas conditions at very large radii
($ > 4 \times 10^{17}$ cm $\approx 190''$). These conditions, however,
cannot
be safely extrapolated from our dust continuum or NH$_3$ data, as they
correspond more to the ambient cloud or envelope than to the dense
core itself, and favoring again simplicity, we have modeled 
them using constant gas parameters determined from an approximate fit to the 
line self absorptions of different molecules. For both L1498 and L1517B,
we have introduced a discontinuity in the density profile at 
$4 \times 10^{17}$ cm, where both cores reach a density of approximately
$3 \times 10^3$ cm$^{-3}$, and we have extended these profiles with 
envelopes of constant density equal to $10^3$ cm$^{-3}$ up to 
a radius of $8 \times 10^{17}$ cm. To match the self absorption 
features, we have set the velocity of this envelope equal to zero 
for L1517B and to 0.35 km s$^{-1}$ for L1498 (inward motion), and the
turbulent component equal to 4 and 5 times the central turbulence, for
L1517B and L1498 respectively. As we will see below, the inward-moving
envelope component in L1498 seems only present in the front part of
the cloud and most likely represents an unrelated foreground cloud,
so we have neglected the contribution of the back side of the
envelope in this object. It should be stressed that our parameterization 
of the cloud envelopes is highly simplified
and it has been carried out with the only purpose of fitting the shape of
the self absorbed lines. These envelopes have no effect on the 
emission of the thin lines used to determine the abundance profiles
of each species, so their role in the modeling is almost cosmetic. Detailed
mapping of the environment surrounding the cores is necessary to
further constrain the envelope properties.

Having fixed the physical parameters of the cores (summarized in Table 3),
the only free parameter left to fit the emission of each species is
the radial profile of its abundance. To determine this profile,
we have searched for Monte Carlo solutions of the radiative transfer 
that fit simultaneously the radial profile of integrated 
intensity and the central spectrum of each observed transition. 
This approach is similar to that of TMCWC, with the  
novelty that the present analysis uses higher 
spatial resolution observations, and that for each 
species we simultaneously model multiple transitions.
This new analysis results in a more detailed and accurate
derivation of the physical and chemical structure of the cores.

\begin{table}
\caption[]{Monte Carlo model fits
\label{tbl-3}}
\[
\begin{array}{lccc}
\hline
\noalign{\smallskip}
\mbox{Parameter} & \mbox{L1498}  & \mbox{L1517B} & \mbox{Note}  \\
\noalign{\smallskip}
\hline
\noalign{\smallskip}
\multicolumn{1}{c} {\mbox{Core}}\\
\hline
\noalign{\smallskip}
r_{max} \mbox{~~(cm)} & 4\times 10^{17} & 4\times 10^{17} & \\
V_{LSR} \mbox{~~(km s$^{-1}$)} & 7.8 & 5.8 & \\
dV/dr\vert_{in} \mbox{~~(km s$^{-1}$ pc$^{-1}$)} & 0 & 0 & (1) \\
dV/dr\vert_{out} \mbox{~~(km s$^{-1}$ pc$^{-1}$)}  & -0.5 & 2.0 & (1) \\
r_{br} \mbox{~~(cm)} & 1.75 \times 10^{17} & 1.5 \times 10^{17} & (1) \\
\Delta V_{NT} \mbox{~~(km s$^{-1}$)} & 0.125 & 0.125 & \\
T_k \mbox{~~(K)} & 10 & 9.5 & (2) \\
\noalign{\smallskip}
\hline
\multicolumn{1}{c} {\mbox{Envelope}}\\
\hline
\noalign{\smallskip}
r_{max} \mbox{~~(cm)} & 8\times 10^{17} & 8\times 10^{17} & \\
V_{LSR} \mbox{~~(km s$^{-1}$)} & 7.45 & 5.8 & \\
\Delta V_{NT} \mbox{~~(km s$^{-1}$)} & 0.625 & 0.500 & \\
\noalign{\smallskip}
\hline
\multicolumn{1}{c} {\mbox{Molecular abundances}}\\
\hline
\noalign{\smallskip}
X_0(\mbox{NH$_3$}) & 1.4 \times 10^{-8} & 1.7 \times 10^{-8} & (3) \\
\beta & 3 & 1 & (4) \\
X_0(\mbox{N$_2$H$^+$}) & 1.7 \times 10^{-10} & 1.5 \times 10^{-10} & (4) \\   
X_0(\mbox{C$^{18}$O}) & 0.5 \times 10^{-7} & 1.5 \times 10^{-7} & (5) \\   
r_{hole}(\mbox{C$^{18}$O}) \mbox{~~(cm)} & 1.5 \times 10^{17} 
& 1.75 \times 10^{17} & (5) \\
X_0(\mbox{CS}) & 3 \times 10^{-9} & 3 \times 10^{-9} & (5) \\   
r_{hole}(\mbox{CS}) \mbox{~~(cm)} & 1.0 \times 10^{17} & 1.15 \times 10^{17} 
& (5) \\
\noalign{\smallskip}
\hline
\end{array}
\]
\begin{list}{}{}
     \item[(1)] $dV/dr = dV/dr\vert_{in}$ if $r<r_{br}$ and 
     $dV/dr = dV/dr\vert_{out}$ if $r>r_{br}$
     \item[(2)] Constant for all $r$
     \item[(3)] para-$X$(NH$_3$) = $X_0$(NH$_3$) $(n(r)/n_0)^\beta$ with 
     $10^{-9}$ threshold
     \item[(4)] Constant in L1517B. For L1498, abundance drops by factor  3
     at $r=1.8\times 10^{17}$ cm
     \item[(5)] $X = X_0$ for $r>r_{hole}$ and $X = 0$ for $r<r_{hole}$
\end{list}
\end{table}

\subsection{NH$_3$: constant temperature and central abundance enhancement}

\begin{figure}
\centering
\resizebox{\hsize}{!}{\includegraphics{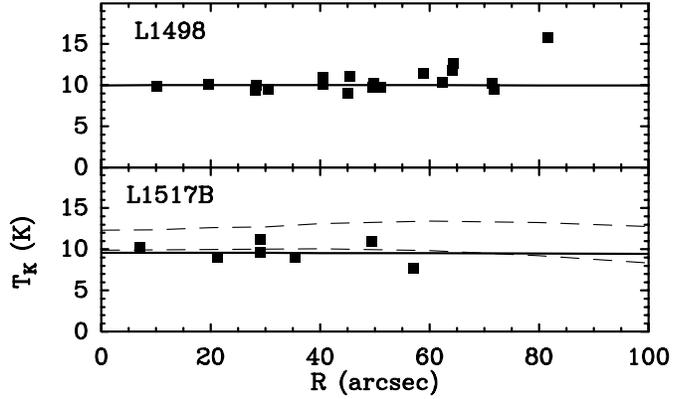}}
\caption{Radial profiles of gas kinetic temperature for L1498 and L1517B 
derived from NH$_3$. The squares represent real data and 
the solid lines
represent the result of applying the same NH$_3$ analysis used
for the data to spectra generated with a Monte Carlo model
having constant temperature (10 K for L1498 and 9.5 K for L1517B).
The dashed lines are the expected NH$_3$-derived $T_k$ profile 
for two temperature equilibrium models (see text for details).
\label{fig4}}
\end{figure}

TMCWC used NH$_3$ data to determine the gas kinetic temperature
of several cores including L1498 and L1517B. Given the importance
of this parameter and the recent
calculations of the expected temperature structure of dense cores
\citep{eva01,zuc01,gal02}, here we 
present an improved NH$_3$ analysis using the new density profiles
of section 3.1 and a more accurate temperature determination method.
Our new method is an updated version of the
classical analysis (e.g., \citealt{wal83}) calibrated
with Monte Carlo radiative transfer models, and as detailed in 
Appendix B, it provides accurate temperature determinations 
for realistic core conditions. With its help, we first
estimate the L1498 and L1517B temperature profiles directly from 
the NH$_3$ data, 
and then use the full  Monte Carlo model to fit the observed 
radial profiles and spectra deriving NH$_3$ abundance profiles
(and demonstrating the self consistency of the temperature determination).

To derive the temperature profiles of L1498 and L1517B, we first 
estimate the $T_R^{21}$ rotation temperature for each observed 
position using the NH$_3$ lines.
This involves fitting the hyperfine (hf) structure of
the NH$_3$(1,1) spectra (where we find that the 
hf components are in LTE), fitting single gaussians
to the NH$_3$(2,2) spectra, and estimating the ratio between the
(2,2) and (1,1) populations (see \citealt{bac87} for details).
Then, we use the analytical expression derived in Appendix B
to convert each rotation temperature into a gas kinetic 
temperature. The resulting temperature radial profiles, 
presented in Fig. 4, show that in each core
the gas kinetic temperature is to very good approximation constant 
with radius and almost equal to 10 K with an rms of about 1 K. 

The temperature profiles in Fig. 4 represent line-of-sight
averages spatially smoothed with a $40''$ gaussian (telescope resolution). To
test the possibility that a realistic non-constant temperature
distribution could mimic the constant temperature profile
of our NH$_3$ analysis, 
we have run a Monte Carlo NH$_3$ model for L1517B using the temperature
law recently calculated by \citet{gal02} (and kindly adapted to L1517B 
by Daniele Galli). The synthetic NH$_3$ spectra from this model have been
analyzed as the real data and used to compute an NH$_3$-derived temperature 
profile. As the upper dashed line in Fig. 4 
shows, a model with the ``standard'' cosmic ray ionization rate of 
$\zeta = 3 \times 10^{-17}$ s$^{-1}$ \citep{gol01}, overestimates the 
core temperature at all radii (in fact, it overestimates the (2,2) intensity by
more than a factor of 2). However, reducing $\zeta$ by a 
factor of 5 (lower dashed curve) produces a temperature profile consistent 
with the observations (such a low $\zeta$ value is also favored by 
the chemical models of the L1544 core in Taurus by \citealt{cas02b}, 
but see 
\citealt{mcc03} for the opposite conclusion in a nearby cloud).
This low $\zeta$ temperature profile is in fact almost 
constant, as it has a maximum-to-minimum variation of less than 
1.5 K in the region of interest. We thus conclude that the gas
kinetic temperature in the L1498 and L1517B cores is constant within
about 10\%, and that this property is consistent a with realistic
treatment of the heating and cooling of dense gas (assuming a low $\zeta$ 
value).

\begin{figure*}
\centering
\resizebox{\hsize}{!}{\includegraphics{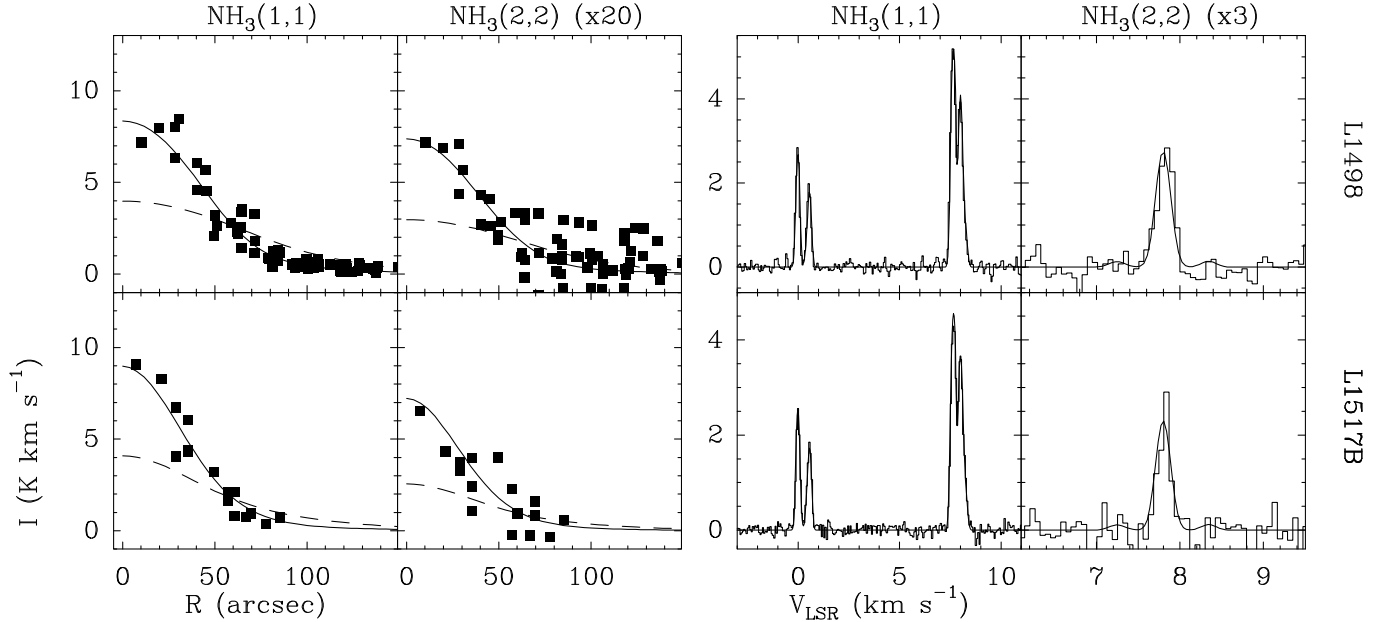}}
\caption{Radial profiles of integrated intensity and central spectra
of NH$_3$(1,1) and (2,2) towards L1498 and L1517B. The squares in the
radial profiles and the histograms in the spectra represent real data.
The solid lines represent the results of Monte Carlo radiative transfer 
models with constant temperature and a central NH$_3$ abundance
enhancement (see text). Note how they simultaneously fit the radial profiles
and the central spectra (they also fit the NH$_3$-derived temperature
profiles, see previous figure). The dashed lines in the radial profiles are
models with constant NH$_3$ abundance, chosen to fit the 
observations at large radius ($\sim 60''$). Note how these flatter profiles
cannot fit the observed central emission.
The velocity scale in L1517B has been shifted by 2 km s$^{-1}$ for
presentation purposes.
(Note that the 1-sigma uncertainty in the
integrated intensity points is of the order of 0.2 K km s$^{-1}$
for NH$_3$(1,1) and 0.8 K km s$^{-1}$ for NH$_3$(2,2) multiplied by 20,
and it is therefore similar to the size of the squares.)
\label{fig5}}
\end{figure*}

Given the above results, we model both L1498 and L1517B as
having constant temperature, and for each core, we search for the NH$_3$ 
abundance profile that fits simultaneously the observed (1,1) and
(2,2) radial profiles of integrated intensity together with the
central spectra of both transitions. We do this by using our full non LTE 
Monte Carlo radiative transfer analysis, and convolve the results with
a $40''$ gaussian to simulate an observation with the Effelsberg antenna
(see Appendix B for details). For L1517B, our best model is identical
to that in TMCWC, as we use the same density profile.
This means that the gas temperature has a constant 
value of 9.5 K and that the para-NH$_3$
abundance increases inwards following the law  
$X(r) = X_0 (n(r)/n_0)^\beta$, where
$X_0 = 1.7 \times 10^{-8}$ and $\beta = 1$
(the above expression is only valid for $X > 10^{-9}$).
Our best fit model for L1498
is slightly different from that in TMCWC because of our improved
density profile and core peak determination from the new continuum 
data. This model has a constant temperature of 10 K and a para-NH$_3$ 
abundance that increases inwards following the same law as in L1517B, 
but this time with $X_0 = 1.4 \times 10^{-8}$ and $\beta = 3$, and 
again with a $10^{-9}$ threshold. 
(Note that if the ortho-to-para ratio of NH$_3$ equals 1,
the total NH$_3$ abundance is 2 times the 
measured para-NH$_3$ abundance.)  
As Fig. 5 shows, these NH$_3$ models fit simultaneously the radial 
profile of integrated intensity and the central
emergent spectra for both NH$_3$(1,1) and (2,2). They also fit
NH$_3$-derived temperature profiles of Fig. 4 (see solid lines).

Although initially unexpected, the higher central NH$_3$ abundance in
L1498 and L1517B is a necessary element of the radiative transfer models. To 
illustrate this point, we also
present in Fig. 5 (dashed lines) the results of constant abundance models
chosen to fit the emission at intermediate radii ($\approx 70''$). As can
be seen in the figure, these models predict radial profiles that are too
flat for both NH$_3$(1,1) and (2,2), and are therefore inconsistent with 
the data. The higher central abundance, thus, is unavoidable, 
and sets apart the NH$_3$ molecule among all other observed species
(see Tafalla et al. 2004, in preparation, for an extensive
chemical survey of L1498 and L1517B). In section 4.2 we discuss the
possible origin of this behavior.

\subsection{N$_2$H$^+$: constant abundance}

\begin{figure*}
\centering
\resizebox{\hsize}{!}{\includegraphics{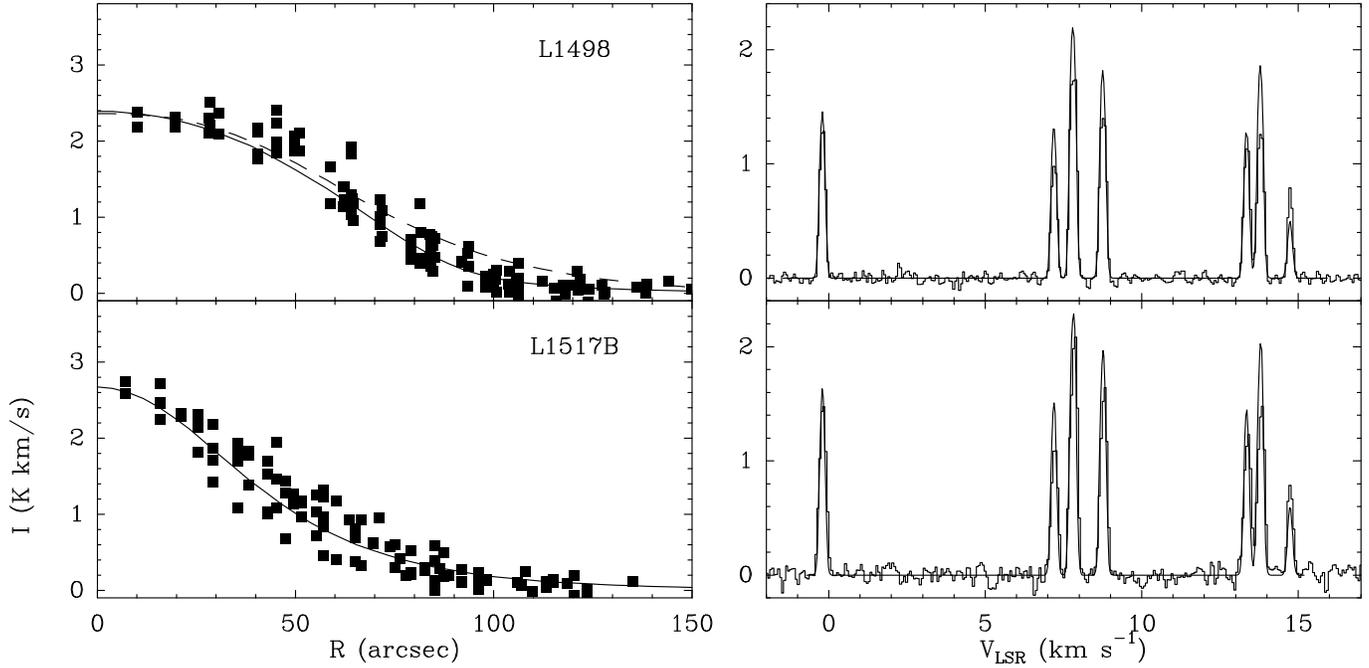}}
\caption{Radial profiles of integrated emission (left) and central emerging 
spectra (right) for N$_2$H$^+$(1--0) toward L1498 and L1517B. Observed
data are represented by squares in the radial profiles (sum of all
hf components) and by histograms in the spectra (main beam brightness 
temperature). The solid lines represent the results from best fit 
Monte Carlo models: for L1517B, the model has a constant
N$_2$H$^+$ abundance of $1.5 \times 10^{-10}$, and for L1498
the model has a constant abundance of $1.7 \times 10^{-10}$
with a drop of a factor of 3 for radii larger than $1.8 \times 10^{17}$ cm.
The dashed line in the radial profile of L1498 shows the expected emission 
for a constant abundance model without outer drop.
The velocity scale in L1517B has been shifted by 2 km s$^{-1}$ for 
presentation purposes.
(Note that the 1-sigma uncertainty in the
integrated intensity points
is of the order of 0.05 K km s$^{-1}$
and it is therefore smaller than the size of the squares.)
\label{fig6}}
\end{figure*}

\begin{figure}
\centering
\resizebox{\hsize}{!}{\includegraphics{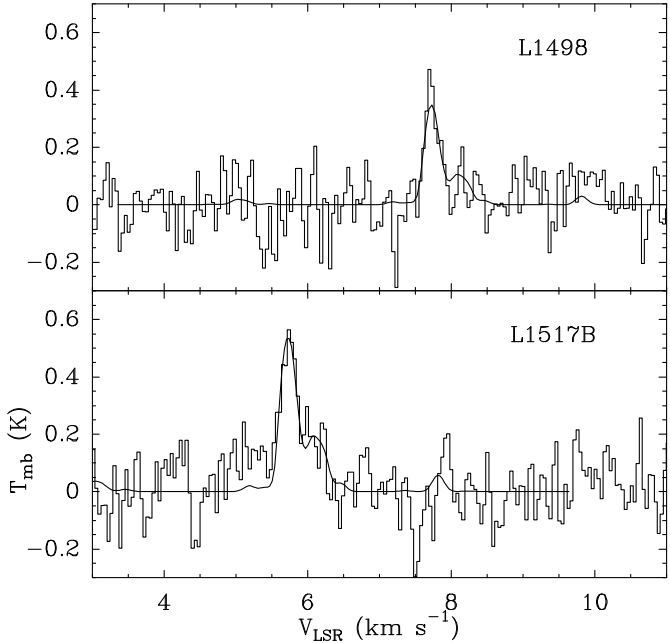}}
\caption{Observed N$_2$H$^+$(3--2) spectra (histograms) and 
predictions (lines) for the same models shown in Figure 6.
The models derived to fit the 1--0 transition also fit the 3--2 spectra.
{(Note that the multiple peaks in the spectra arise from hyperfine structure, and 
do not represent multiple velocity components.)}
\label{fig7}}
\end{figure}

Like NH$_3$, N$_2$H$^+$ presents centrally peaked distributions in 
our maps of Fig. 3, and this suggests that the N$_2$H$^+$ abundance
does not vary greatly over the core interiors. 
In this section we quantify this impression
by solving the N$_2$H$^+$ radiative transfer with our Monte Carlo code.
This code now needs to take
into account the hyperfine structure introduced by the two N atoms, 
because (in contrast with the almost thermalized NH$_3$) the N$_2$H$^+$ 
hyperfine structure affects the line trapping and the level excitation.
Unfortunately, an exact treatment of the N$_2$H$^+$ radiation transfer
with hf structure is not yet possible for two reasons. First, 
the collision rates between the individual hf components are not known,
and only approximate hf-averaged values are available (see below). Second, the
hf structure introduces the possibility
of partial overlap between lines, complicating the problem. 

TMCWC
presented a simplified solution of the N$_2$H$^+$ transfer which ignored 
the hf structure in the excitation calculation but considered it in the
computation of the emergent spectrum. As mentioned there, the main
problem with this approach is that it neglects the decrease in trapping
caused by the splitting. Here, to take this effect into account, we improve 
the model  and treat explicitly the hf splitting in the low J transitions
(up to J=2), where the trapping decrease is expected to be important,
while we assume the rest of transitions are degenerate (up to J=6). This
improved approach requires increasing the number of levels considered in 
the Monte Carlo code to 23 and the number of transitions to 44 (no
line overlaps are considered). For collision rates, we 
use the HCO$^+$-H$_2$ rates from \citet{flo99} and derive the individual
coefficients between N$_2$H$^+$ hf components by assuming that 
collisions do not distinguish between hf levels. By using these
simplified collision rates and neglecting line overlaps we lose the 
ability to reproduce the non LTE excitation known to occur in the
J=1--0 line \citep{cas95}. This excitation anomaly, however, seems to be a 
10\% effect in the level population, and therefore is of little 
consequence for our abundance estimate. 

To better constrain the N$_2$H$^+$ abundance profiles, we fit simultaneously
the N$_2$H$^+$(1--0) radial profile of integrated intensity and the
emergent spectra toward the core center for both N$_2$H$^+$(1--0)
and N$_2$H$^+$(3--2). We start our analysis by exploring models with
constant abundance, which we find fit the data rather well.
As illustrated in Fig. 6 and 7, a constant abundance model with X(N$_2$H$^+$) = 
$1.5 \times 10^{-10}$ fits automatically all L1517B observations
and no additional change is required in the abundance profile 
to fit this source. (Note that our model spectra are slightly brighter 
than the observations despite having the same integrated area because the
real N$_2$H$^+$ lines seem to have additional broadening, most likely
caused by unresolved hyperfine structure, see section 3.7.1). For
L1498, a constant abundance model with X(N$_2$H$^+$) = $1.7 \times 10^{-10}$
fits well both the J=1--0 and 3--2 spectra (not shown) and
almost fits the J=1--0 radial profile, although it tends to 
overestimate the intensity at large radius (dashed lines in Fig. 6).
To improve this fit, we have modified the constant abundance model 
by introducing an abundance drop of a factor of 3 for
radii larger than $1.8 \times 10^{17}$ cm ($85''$ at 140 pc). This 
new model fits all L1498 data rather well, as can be seen by the
solid lines in Fig. 6 and 7 (the slight deviation from flat top 
at the center of L1498 probably arises from a similar deviation in our density 
profile, see Fig. 2). 

In contrast with our conclusion that the N$_2$H$^+$ abundance is
constant or close-to-constant 
at the centers of L1498 and L1517B, \citet{ber02} find
a factor of 2 drop in the N$_2$H$^+$ abundance 
at the center of the B68 core. Such a significant drop seems excluded
in L1517B or L1498, not just from our modeling but from the
lack of any significant drop in the N$_2$H$^+$ emission near the
continuum peak (see maps in Fig. 3).
This contrast with the behavior in B68 is especially significant
in the case of L1517B, as both B68 and L1517B
have central densities of about $2 \times 10^5$ cm$^{-3}$ and are 
well fitted by isothermal models (\citealt{alv01}, our section 3.1).
Given the importance of N$_2$H$^+$ as a tracer of dense gas, further
work should be carried out to understand the discrepancy between
these cores.

\subsection{CO isotopomers: central depletion}

\begin{figure*}
\centering
\resizebox{15cm}{!}{\includegraphics{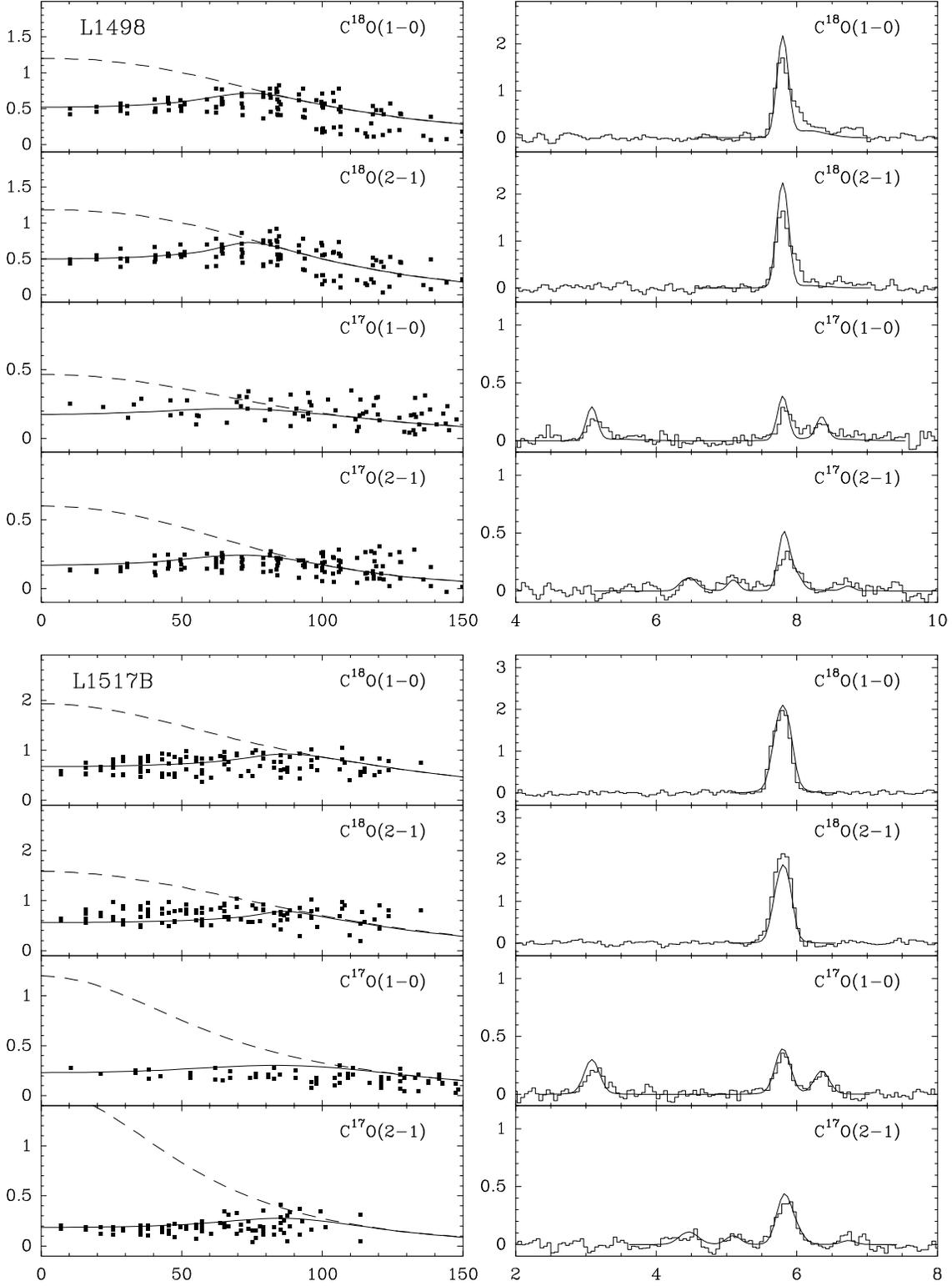}}
\caption{Radial profiles of integrated emission (left) and central emerging
spectra (right) for C$^{18}$O(1--0), C$^{18}$O(2--1), C$^{17}$O(1--0), 
and C$^{17}$O(2--1) toward L1498 and L1517B. Observed
data are represented by squares in the radial profiles 
and by histograms in the spectra (main beam brightness
temperature). The dashed lines represent constant abundance models
chosen to fit the emission at radius $100''$
(X(C$^{18}$O)$=0.5 \times 10^{-7}$ in L1498 and $1.5 \times 10^{-7}$
in L1517B) and, as the figure shows, 
they fail to fit the central emission by a large margin.
The solid lines are models with the same abundance value for large radii but
having a central hole with no molecules ($r_{hole}=71''$ for L1498 and 
$83''$ for L1517B). These models fit simultaneously all transitions 
at all radii. A C$^{18}$O/C$^{17}$O isotopic ratio
of 3.65 is assumed in both cores. (Note that the 1-sigma uncertainty 
in the integrated intensity points is of the order of 0.03 K km s$^{-1}$ 
and it is therefore similar to the size of the squares.)
\label{fig8}}
\end{figure*}

The central holes in the CO isotopomer maps of Fig. 3 indicate graphically
that the abundance of these species drops systematically toward
the peak of both L1498 and L1517B. To determine 
quantitatively these abundance drops, we use again 
the Monte Carlo radiative transfer model to fit the observations
with a variable abundance profile
(for further details on the Monte Carlo modeling, see TMCWC).
To better constrain the abundances
and to provide a self consistency check to our calculations, we use 
simultaneously all the CO isotopomer data we have available. Thus,
in addition to fitting the C$^{18}$O(1--0), C$^{18}$O(2--1), 
and C$^{17}$O(2--1) data from the IRAM 30m telescope, we use the 
lower angular resolution C$^{17}$O(1--0) data from FCRAO already
presented in TMCWC. Also, taking into account previous
work on isotopic abundances in the ISM, we assume a constant
C$^{18}$O/C$^{17}$O abundance ratio of 3.65 \citep{pen81}. In this way,
our determination of the CO isotopomer abundances reduces for each core
to finding a single abundance profile that fits simultaneously
four radial profiles of integrated intensity and four emergent spectra.

Before attempting to reproduce the observations, we test  
the conclusion that constant abundance models cannot fit 
the data from neither L1498 nor L1517B. To do this, we fix the
C$^{18}$O abundance so that the models fit the observations at a radius
of $100''$, and compare the model predictions with the 
observations at smaller radii. As the dashed lines in the radial profiles
of Fig. 8 show, the constant abundance models fail to fit the data for all
CO isotopomers in both L1498 and L1517B. They do so by a wide margin 
(more than a factor of 2), and this proves that constant abundance 
C$^{18}$O/C$^{17}$O models are inconsistent with observations.
A sharp drop of abundance toward the center is therefore required.

To model the CO central abundance drop, we have explored different 
abundance profiles, starting with the exponential forms introduced
by TMCWC ($X \sim \exp(-n(r))$). Our higher angular resolution 
data, however, suggest that these forms do not provide
a fast enough abundance drop to fit the integrated
intensity near the core centers, and that faster decreases 
are needed to reproduce the data (note that because of the central 
flattening of the density profiles, the exponential forms do not
necessarily result in dramatic abundance drops). Thus, we have explored
alternative abundance profiles, like squares of the exponential forms
or central holes. Unfortunately, if the abundance toward the core
center is low enough that the emission is dominated by the outer layers
(order of magnitude drop), 
the exact form of the abundance profile is not very critical. 
Given this degeneracy of the solution, we have
chosen the simplest form, a (step) central hole, and 
used it as basic model for the CO (and CS) abundance profiles 
in L1498 and L1517B.
Although in some cases a slightly different abundance profile may produce
a better fit to the integrated intensity 
radial profile (as in the case of CS in L1517B, see
below), we have preferred for consistency and inter-comparability to fit
all the abundance drops with central holes. 

Fig 8 shows a comparison between the central hole models (solid lines)
and the observations for both L1498 and L1517B. These models are 
characterized by
just two parameters (external abundance and hole radius, see Table 3)
and provide a simultaneous fit to the four integrated intensity radial
profiles and the four central spectra of C$^{18}$O and C$^{17}$O.
Their ability to fit the data shows that the central abundance drop
in each core is sudden and deep, and that the data are consistent with
both L1498 and L1517B having no CO molecules in their innermost 
region $(r<70'')$. Although the radius of the
central hole is not uniquely determined by the observations (it is
slightly coupled to the initial abundance estimate), the true values cannot
be very far from our estimates. Thus, we
combine the radius of the abundance drop (Table 3) with the 
density profiles (Table 2) in order to estimate that the density at which
CO molecules disappear from the gas phase is
$7.8 \times 10^4$ cm$^{-3}$ for L1498 and
$2.5 \times 10^4$ cm$^{-3}$ for L1517B. 

\subsubsection{Deviations from spherical symmetry}

Our radiative transfer analysis has assumed spherical symmetry, and 
our Monte Carlo models have been used to fit azimuthal averages of the observed
emission. This means that our radial profiles with a central abundance drop 
represent spherical averages of the distribution of CO molecules in each core. 
As the maps in Fig. 3 show, however, 
the distribution of CO emission
in both L1517B and L1498 is not fully circular (or elliptical), and superposed
upon the central emission drop, there is a pattern of secondary peaks
and valleys. In L1517B, for example, the CO emission is brighter toward 
the west, and in L1498 there are bright spots toward the southeast and the west.
The asymmetries in the CO emission appear in all transitions but
do not have counterparts
in the continuum or NH$_3$ emission (Fig. 3). This suggests that they 
do not arise from
asymmetries in the distribution of density or temperature, but that they result
from true non-azimuthal asymmetries in the distribution of CO abundance. 
The detailed modeling of these asymmetries is outside the scope
of our spherically symmetric radiative transfer analysis, but a simple
inspection of the maps in Fig. 3 suggests that the abundance variations
needed to produce them
are at the level of a factor of two. These variations, therefore,
are only perturbations on the (order of magnitude) drop 
in abundance at high densities, 
but as we will see in section 4.3, they offer interesting clues concerning
the process of core formation.

\subsection{CS: central depletion}

\begin{figure*}
\centering
\resizebox{15cm}{!}{\includegraphics{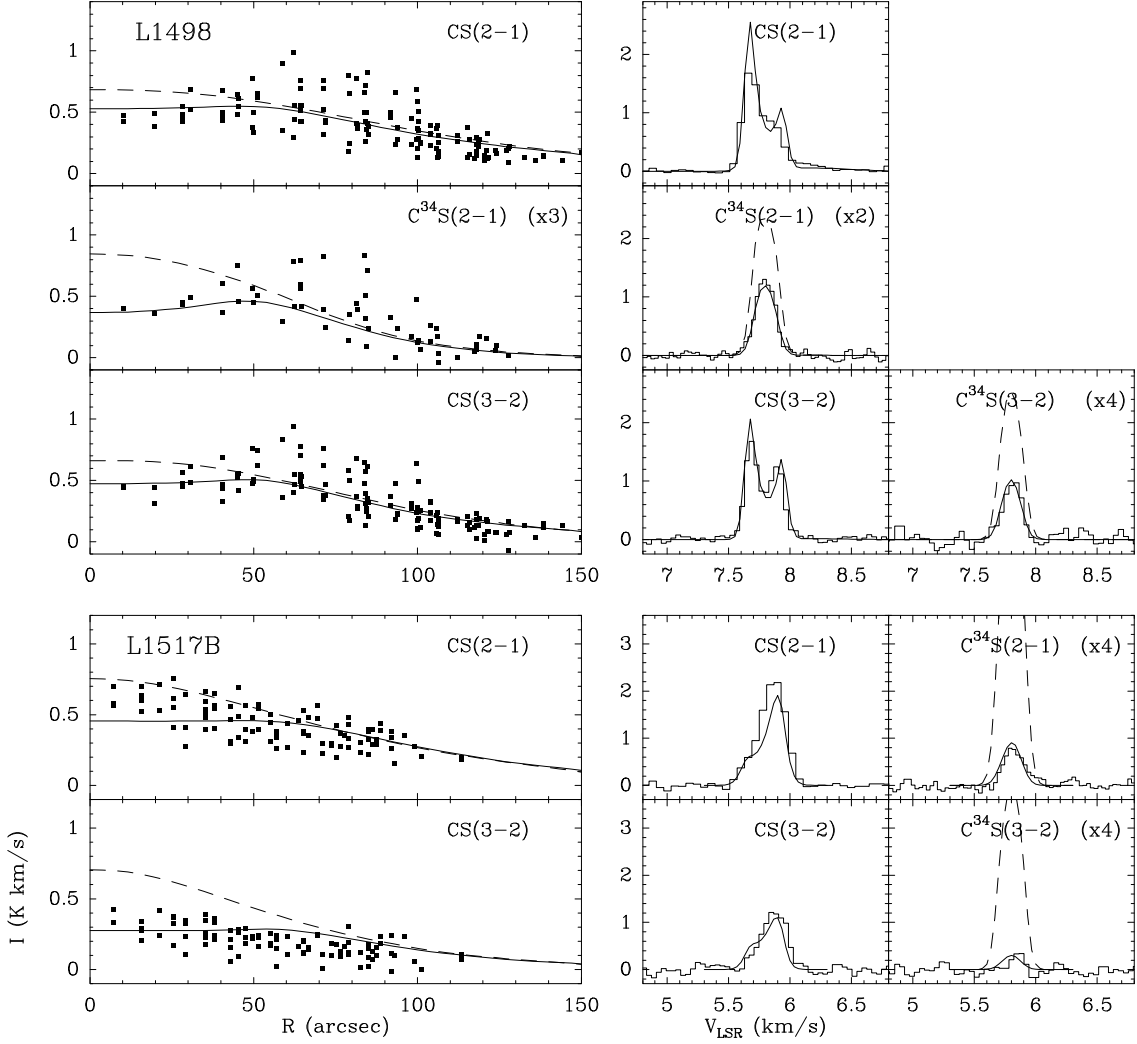}}
\caption{Radial profiles of integrated emission (left) and central emerging
spectrum (right) for CS(2--1), CS(3--2), C$^{34}$S(2--1),
and C$^{34}$S(3--2) toward L1498 and L1517B. Observational
data are represented by squares in the radial profiles
and by histograms in the spectra (main beam brightness
temperature).  The dashed lines represent constant abundance models
chosen to fit the emission at radius $100''$
(X(CS)$=3 \times 10^{-9}$ in both L1498 and 
L1517B) and, as the figure shows,
they fail to fit the central emission by a large margin.
The solid lines are models with the same abundance value for large radii but
having a central hole with no molecules ($r_{hole}=48''$ for L1498 and
$55''$ for L1517B). These models fit simultaneously all transitions
at all radii. A CS/C$^{34}$S isotopic ratio
of 22.7 is assumed in both cores. (Note that the 1-sigma uncertainty 
in the integrated intensity points
is of the order of 0.03 K km s$^{-1}$
and it is therefore similar to the size of the squares.)
\label{fig9}}
\end{figure*}

We finish our abundance analysis studying the distribution of
the CS molecule. Unfortunately, its emission is more difficult to model 
because the CS lines from both cores are very optically thick and 
strongly self absorbed. This 
can be seen in Fig. 9, where the main CS isotopomer lines appear
asymmetric and double-peaked while the spectra of the rare C$^{34}$S
species are single gaussians. These self absorbed spectra depend strongly 
on the physical parameters of the low-excitation outermost layers 
that give rise to the absorption ($r>0.1$~pc) and for which we have very 
little information in terms of density and temperature. Thus, although 
the Monte Carlo code can treat accurately the transfer of radiation
under optically thick conditions, our CS modeling is limited by 
our ignorance of the gas properties in the outer core regions.
If these properties deviate 
from the extrapolation of our simple models, the radiative transfer
solution for the main isotopomer of CS is compromised. To minimize this
effect, we have complemented our CS observations with C$^{34}$S data.

As with CO, we test the CS models simultaneously against our full CS
isotopomer data set, which consists of maps and spectra of CS(2--1) and
CS(3--2), and central spectra of C$^{34}$S(2--1) and C$^{34}$S(3--2)
(plus a full C$^{34}$S(2--1) map of L1498). To minimize the
number of free parameters, we set the
CS/C$^{34}$S isotopic ratio to its solar value of 22.7, since this number
seems consistent with ISM determinations \citep{luc98}. An acceptable
abundance curve, therefore, must fit simultaneously the 4 emergent spectra
and 2 radial profiles (3 in L1498) shown in Figure 9.

Our first set of Monte Carlo models have spatially constant CS abundance,
set to a value that fits the integrated intensity radial profile 
at a radius of $100 ''$. As the dashed lines in Fig. 9 show, these
constant abundance models predict a significant concentration of the emission 
toward the core center, which is not seen in the data. This is 
more evident in L1517B because of its higher central density concentration,
but is also clear in the flatter and more CS-opaque L1498 core,
especially in the
thinner C$^{34}$S(2--1) line, where the central deviation between model
and data is of at least a factor of two. The failure of the constant 
abundance models to fit the data proves that the CS abundance must
drop toward the centers of both L1498 or L1517B
(in agreement with previous lower resolution results, see 
\citealt{kui96} and TMCWC).

To model the central decrease of the CS abundance, 
we have again explored different families of abundance profiles.
Models with a central hole, for example, seem slightly 
superior to exponential drop models 
in the case of L1498, although both models fit the data in L1517B.
Unfortunately, a detailed comparison between the different 
forms of the central abundance drop is difficult
because of the dependence of the results
on the CS self absorption, and for this reason, we only present the results
of models with a central hole. These models provide reasonable
fits to the observations, as illustrated in Fig. 9, 
which shows that central hole models (solid lines)
match simultaneously for each core all radial
profiles of integrated intensity and all central spectra.

The simultaneous fit of both the self absorbed CS lines and the gaussian 
C$^{34}$S spectra is a critical test for the models. 
The CS self absorption arises as a natural consequence 
of the combination of high optical depth and subthermal excitation 
in the model low-density outer layers, and this makes this
feature the most sensitive
tracer of the velocity field in the outer core (as mentioned in 
section 3.2).  The red shifted CS selfabsorption in L1498 indicates a 
contraction motion, while the blue shifted self absorption in L1517B
requires expansion (in agreement with
previous work, see \citealt{kui96,lee99b}, and TMCWC).
These velocity gradients, however, do not continue all the way to 
the core center, since otherwise the central C$^{34}$S spectra would 
not have the same linewdith (within the noise) as the N$_2$H$^+$ and 
NH$_3$ lines. Thus, we have parametrized the velocity field 
as constant up to a 
radius of $1.75 \times 10^{17}$ cm ($83''$) in L1498 and 
$1.5 \times 10^{17}$ cm ($71''$) in L1517B, and then as increasing/decreasing 
linearly with radius for larger distances (Table 3). 
To fit the broader CS(2--1) self absorption and the slight wing in L1498, 
we have additionally
required that the CS abundance increases by a factor of 5 
in the outer envelope ($r>4 \times 10^{17}$ cm $\approx 190''$). 
As mentioned in section 3.2, however, this
envelope most likely represents 
a foreground red component that also gives rise to a
red wing in the C$^{18}$O spectra (Fig. 8, see also \citealt{lem95}).
Foreground contamination, in fact,  may also be the cause of 
the blue shifted self absorption in L1517B, as the nearby L1517A core
is slightly blue shifted with respect to L1517B \citep{ben89}. Thus, the
velocity gradients traced with the CS self absorption most likely reflect the 
conditions in the region where the cores meet the extended ambient
cloud ($r>0.1$ pc) and not an intrinsic core velocity pattern.
Although these motions affect strongly the shape of the
CS spectra, they seem to involve only the core outermost layers and 
not the dense (probably star-forming) part.
Their interpretation as global core contraction 
or expansion patterns seems therefore not warranted.

The reasonable fit provided by the central-hole models shows that the 
CS abundance drop in L1498 and L1517B is as dramatic as that of CO.
This implies that the centers of these two cores may
lack CS molecules for radii smaller than $48''$ in L1498
and $55''$ in L1517B (Table 3), and that the CS emission in these systems 
only traces the outer core layers. As with CO, we can convert the 
radius of the central hole into a critical density above which
CS molecules disappear from the gas phase. Using the density profiles
of Section 3.1, we derive a maximum density of $8 \times 10^4$ cm$^{-3}$
for L1498 and $4 \times 10^4$ cm$^{-3}$ in L1517B, which are similar to
but slightly larger than those measured for the CO isotopomers. 
Thus, CS seems to freeze-out from the gas phase at slightly larger
densities than CO, and therefore traces slightly more central 
parts of cores. The innermost core regions, however, are totally 
invisible in CS, even when observed with optically thin C$^{34}$S emission.

A look at the maps of Fig. 3 shows that the CS emission presents the
same azimuthal asymmetries found in CO: relative maxima
toward the SE and W in L1498 and toward the W in L1517B. This similarity
with the pattern of C$^{18}$O asymmetries and the fact 
that in L1498 the asymmetry
can also be seen in the thinner C$^{34}$S(2--1) emission strongly
suggest that these features are real and not the result
of self absorption effects. Thus, following
the arguments presented in the study 
of CO (section 3.5.1), we conclude that in each core the distribution
of CS abundance is not azimuthally symmetric.
As in CO, this azimuthal asymmetry probably arises from
changes in the abundance of a factor of a few and is superposed upon the
order-of-magnitude central abundance drop. A full discussion
of its possible origin in terms of core formation history is
deferred to section 4.3.

\subsection{Gas kinematics}

Our modeling of the line profiles toward the center of L1498 and L1517B
has sampled the gas velocity field along the central line of sight of 
each core. This field has been found to have a turbulent component 
approximately constant with radius, with a possible increase in the 
outer part of L1498 (but most likely due to another velocity
component). The line-of-sight systemic velocity seems also constant 
over the center of the cores, although it presents gradients
in the outer layers: contraction in L1498 and expansion in L1517B. 
In this section, we turn our attention to the 2-dimensional (plane of the sky) 
velocity field as a complement to the analysis of the 
central line-of-sight. To do this, 
we make gaussian fits to the spectra at all
positions and we study the spatial variations of the fit results.
As only N$_2$H$^+$ and NH$_3$ are
sensitive to the core dense gas, we only use these
species in our velocity study. Both N$_2$H$^+$ and NH$_3$ have the
additional advantage of having hyperfine structure, which 
allows for correction of optical depth effects in the linewidth 
determination.

\subsubsection{Gas turbulence}

\begin{figure}
\centering
\resizebox{\hsize}{!}{\includegraphics{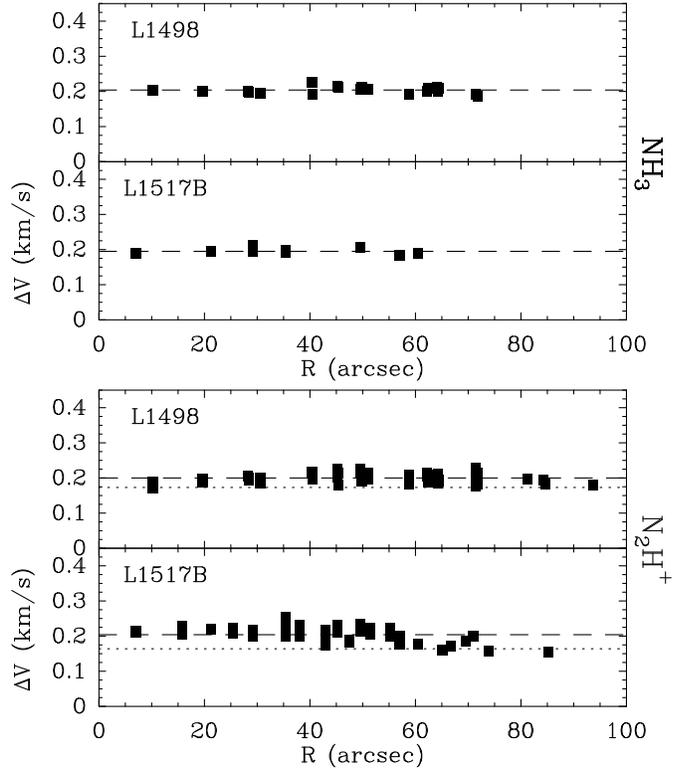}}
\caption{Radial profiles of intrinsic linewidth 
for L1498 and L1517B from NH$_3$
and N$_2$H$^+$ hyperfine-structure fits. 
Note the lack of systematic variations and the extremely low dispersion.
The long-dashed horizontal lines indicate the average value for each
core and molecule (very close to 0.20 km s$^{-1}$ in all cases). The
short-dashed lines indicate the N$_2$H$^+$
linewidth expected when adding the nonthermal component derived from 
NH$_3$ to a thermal N$_2$H$^+$ component at 10 K (see text for details).
To maximize sensitivity, only points with integrated intensity larger than
1.3, 1, 0.6, and 0.6 K km s$^{-1}$ are presented for L1498 NH$_3$, 
L1517B NH$_3$, L1498 N$_2$H$^+$, and L1517B N$_2$H$^+$, respectively.
(As indicated in the text, the 1-sigma uncertainty of the
linewidth estimates, both from NH$_3$ and N$_2$H$^+$, is of
the order of 0.01 km s$^{-1}$,
and it is therefore smaller than the size of the squares.)
\label{fig10}}
\end{figure}

We study the spatial distribution of the gas turbulent component 
using the intrinsic linewidths derived from the hf fits
and determining their radial profiles as we have done before with the 
integrated line intensities. As the intrinsic linewidth is the sum
in quadrature of a thermal and a non thermal component (e.g., 
\citealt{mye83}), and the thermal part is constant because the
gas kinetic temperature is constant (section 3.3), 
any spatial variation of the intrinsic
linewidth has to result from an equivalent spatial variation
of the non-thermal (turbulent) part.

Fig. 10 presents the radial profiles of the NH$_3$(1,1) and N$_2$H$^+$(1--0) 
intrinsic linewidths $\Delta V$ for both L1498 and L1517B.  A simple 
inspection of these profiles suggests that there is no significant 
trend of increase or decrease of linewidth as a function of radius,
meaning that any non thermal component is also spatially constant over
both cores.
To test for the presence of a hidden slope in the linewidth distribution, 
we have fitted the points in Fig. 10 with a function of the form
$\Delta V = a + bR$, where R is the radius in arcseconds and 
$a$ and $b$ are two free parameters. 
In three out of four cases the $b$ coefficient is smaller
than its $rms$, hence the fit is consistent with $b=0$
and the distribution of $\Delta V$ is independent of radius.
In the fourth case (L1517B N$_2$H$^+$ data), the slope is negative, and  
the linewidth slightly decreases with radius (only by 30\% in
$100''$). This slight decrease, which can be seen in the data at large radii, 
is most likely due to errors in the N$_2$H$^+$ hfs fit (see below)
and has no counterpart in the  NH$_3$ linewidth distribution. 
We therefore conclude that the non thermal component of the velocity field
in both L1498 and L1517B is constant with radius for as far as we can measure.

The finding of constant linewidth in L1498 and L1517B agrees
with the results of \citet{bar98}, who found the
same behavior in the interior of four other dense
cores. These authors, however, suggested that linewidths may increase
at the largest observed radii, a behavior not seen in our data.
Even if we include the low sensitivity points at very large radii, the
constant linewidth trend of Fig. 10 remains unchanged, and the scatter due to 
noise dominates the plots.
We therefore conclude that the linewidth does not increase with radius,
or if it does, it starts increasing at a larger radius than suggested by 
\citet{bar98}. 

A puzzling result from our hf analysis is that the 
N$_2$H$^+$(1--0) and NH$_3$(1,1) linewidths are almost equal
(very close to 0.20 km s$^{-1}$). This is unexpected because 
the N$_2$H$^+$ molecule is 1.7 times heavier than NH$_3$ and thus 
its thermal
component is 1.3 times smaller. If both molecules are tracing the same
gas, their turbulent components should be equal, and the lighter NH$_3$
should always present broader lines. To quantify this problem, we have 
assumed that the NH$_3$ hf fit returns the true linewidth (thermal plus 
turbulent), and we have derived its nonthermal part by
subtracting a thermal component at 10 K. When this nonthermal component
is added to the N$_2$H$^+$ thermal part, we find that the N$_2$H$^+$ 
linewidths derived from the hf fit are approximately 20\% larger than 
expected, as illustrated in Fig. 10, where the dotted 
line is the expected N$_2$H$^+$ linewidth. 

The origin of the larger N$_2$H$^+$ non-thermal component is unclear. 
Different linewidths for ions and neutrals are expected
in the presence of magnetic fields, but the observed trend
in both L1498 and L1517B (larger non-thermal component in the ion line) 
is opposite to what is expected theoretically \citep{hou00}.
A more likely explanation is that the N$_2$H$^+$ hf fit overestimates 
the linewidth by about 20\%. This is so because the hf analysis 
of both molecules assumes
LTE among hf components, but this is only correct for the NH$_3$ lines.
N$_2$H$^+$ is notorious for showing non-LTE excitation among its
components \citep{cas95}, and this violates a main assumption of the
analysis. In addition, as our data show and was first noticed by 
\citet{cas95}, there is a discrepancy between the N$_2$H$^+$ linewidth 
derived from the hf fit and the linewidth measured by a Gaussian fit to 
the thinnest (reddest) component. This component appears slightly (5-10 \%) 
narrower than the hf-fit linewidth, contrary to what would be expected
if the hf fit truly corrected for optical
depth broadening. This suggests that N$_2$H$^+$ linewidths derived from
standard hf fits may be slightly overestimated (20\%), either because
of a failure in the LTE analysis or because of the presence of additional
hyperfine splitting due to magnetic spin-rotation interaction of
the H atom (the latter option seems less likely according to recent
estimates of the splitting, Luca Dore private communication).  
Of course, this suggestion does not contradict our conclusion that the
linewidth is constant in L1498 and L1517B, as it is confirmed independently 
(with even less scatter) by the more reliable NH$_3$ analysis (Fig. 10).
According to this analysis, the mean intrinsic (NH$_3$) linewidth is
0.204 km s$^{-1}$ in L1498 and 0.196 km s$^{-1}$ in L1517. Subtracting
thermal components of 10 K for L1498 and 9.5 K for L1517B (section 3.3),
we estimate non thermal turbulent (FWHM) linewidths of 0.121 km s$^{-1}$ 
in L1498 and 0.113 km s$^{-1}$ in L1517B (note that our Monte Carlo
models use 0.125 km s$^{-1}$ as a compromise to fit both NH$_3$
and the broader N$_2$H$^+$).

Even if the turbulent linewidth is constant with radius on average, it
may still present random variations 
across each core, and to explore this possibility we
compare the measured $rms (\Delta V)$ of N$_2$H$^+$ and NH$_3$
($\approx 10^{-2}$ km s$^{-1}$)
with the value expected from the noise level in the
data. To calculate this expected $rms$, we have created at each L1498 and 
L1517B position a pair of simulated
N$_2$H$^+$ and NH$_3$ spectra with the correct intensity and noise level,
but with all positions having the same linewidth. These 
simulated data result from averaging all spectra of each molecule 
and core, creating a high S/N spectrum which has then been scaled
for each position to match the observed integrated intensity,
with noise added to match the observed noise level
(the averaging process broadens the line due to velocity fields, 
but by less than 10\%). This artificial
data set has then been subject to the same hf analysis as the real data,
so its $rms (\Delta V)$ provides a direct measure of the effect of noise
on the measured linewidth. Comparing this $rms (\Delta V)$ with that
measured from the real data, we find that 
$rms (\Delta V)$ of the real data
is $1.1 \pm 0.3,$ times the $rms (\Delta V)$ of the simulated
data, which means that the observed dispersion is consistent with
noise, and that there is no evidence for an additional
source of linewidth variations. This result, which agrees
with the line-of-sight analysis of the Monte Carlo modeling,
indicates that the data are consistent with a single $\Delta V$
value in the inner core ($r<0.05$ pc) within better than 
0.01 km s$^{-1}$ (5\%). 

\subsubsection{Velocity gradients}

\begin{figure}
\centering
\resizebox{\hsize}{!}{\includegraphics{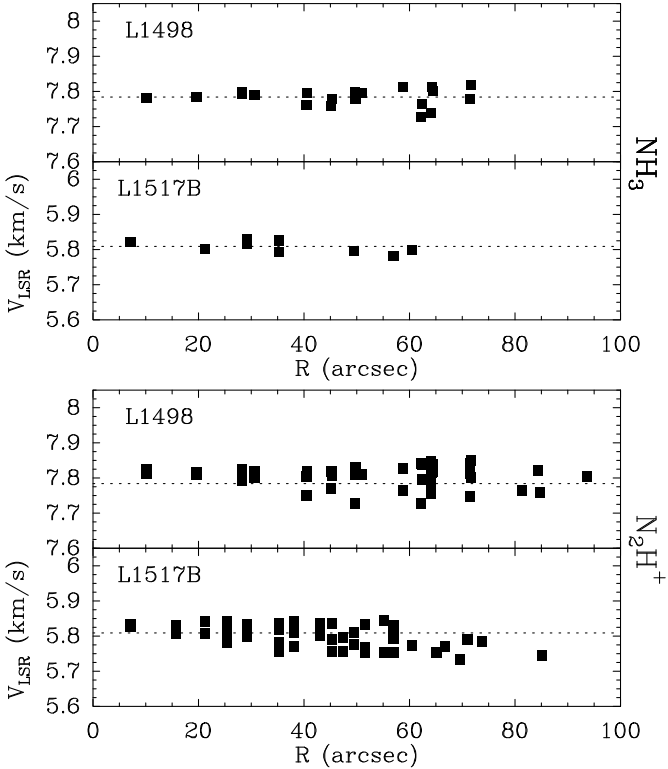}}
\caption{Radial profiles of line center velocity
for L1498 and L1517B derived from NH$_3$ 
and N$_2$H$^+$ hyperfine-structure fits.
The dispersion of the points indicates the presence of bulk 
internal motions. Intensity threshold as in Fig. 10.
(As indicated in the text, the 1-sigma uncertainty of the
velocity estimates, both from NH$_3$ and N$_2$H$^+$, is of
the order of 0.005 km s$^{-1}$,
and it is therefore smaller than the size of the squares.)
\label{fig11}}
\end{figure}

The other kinematic parameter derived from the hf fit of the 
NH$_3$ and N$_2$H$^+$ spectra is
the line center velocity. As with the linewidth, we have
made radial profiles of the line center velocity for
both NH$_3$ and N$_2$H$^+$ in L1498 and L1517B, and they are
presented in Fig. 11. These radial profiles are flat on average, 
and their scatter around 
the mean is larger than the scatter in linewidth by a about a 
factor of 2. This larger scatter of the line center velocity 
is significant according to the model
of simulated spectra used in the previous section, because 
the hf analysis is more sensitive to the line center velocity
than to the linewidth, especially for N$_2$H$^+$. Comparing the data
with the model, we estimate that the observed
scatter in the line center velocity is 0.03 km s$^{-1}$, or
$5.4\pm 2.5$ times larger
than it should be if all positions had exactly the same velocity. This
implies that the $V_{LSR}$ variations seen in the radial profiles of Fig. 11
correspond to real changes in the gas velocity across the core
and not just to noise.

\begin{figure}
\centering
\resizebox{\hsize}{!}{\includegraphics{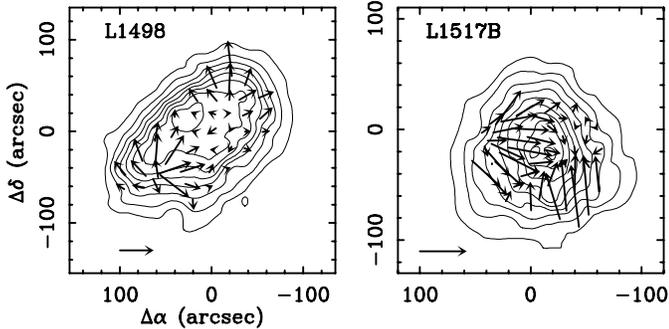}}
\caption{
Gradients of $V_{LSR}$ from N$_2$H$^+$ hf fits. The arrows indicate 
both the direction (from low to high $V_{LSR}$) and intensity of the
gradient (the arrows in the bottom left corners represent 
gradients of 3 km s$^{-1}$ pc$^{-1}$). The contours are N$_2$H$^+$
integrated intensity maps.
\label{fig12}}
\end{figure}

To study the origin of the velocity changes, we first determine their
spatial distribution. Using a program similar to that of \citet{cas02a},  
we have calculated at
each map position the local velocity gradient, both in modulus and
direction, by comparing the $V_{LSR}$ velocity of each point with that
of its neighbors.  A graphical representation of the N$_2$H$^+$ 
results is shown in Fig. 12, and similar plots, but with fewer points, 
were derived for NH$_3$ (their similarity confirms that the velocity 
fluctuations represent real velocity variations). As these maps
show, both the direction and modulus of the velocity gradient
change across both L1498 and L1517B, indicating that the gradients do
not originate from simple global velocity patterns like rotation. This lack
of global patterns, however, does not imply that the velocity 
variations are completely random, because the gradients
are correlated over areas larger than the telescope beam ($26''$)
or the area used to measure individual velocity gradients ($40''$). They
therefore have to arise from bulk gas motions that affect
sizeable fractions of the core material. 

As a first estimate of the importance of 
these motions, we calculate the average
velocity gradient for each core. Both N$_2$H$^+$ and NH$_3$
give consistent results, especially for the position
angle (-20 and -75 degrees for L1498 and L1517B, respectively).
For the modulus of the gradient, the NH$_3$ data give a slightly
lower value (10\% for L1498 and 40\% for L1517B), so we average 
the N$_2$H$^+$ and  NH$_3$ values and derive
0.75 and 1.1 km s$^{-1}$ pc$^{-1}$ for L1498 and L1517B, respectively.
These values are similar to those found by \citet{goo93} in 
other cores, again an indication that L1498 and L1517B are 
representative Taurus cores. Our value for L1498, in addition, 
is in excellent agreement with a similar estimate by \citet{cas02a}.
The gradients
measured in this manner, however, are strict lower limits, as in  
addition to the average across the sky required to estimate
the mean gradient, the velocity estimate at each position already
represents an average along the line of sight. This average necessarily
cancels spatial variations of the velocity, given the lack 
of coherence over scales of a core diameter (Fig. 12). Unfortunately, this
effect cannot be corrected without a full model of the core kinematics, and
as a simple alternative for estimating the velocity 
differences between parts of the core, we measure the spread of 
$V_{LSR}$ in Fig. 11, which is approximately 0.1 km s$^{-1}$
($=3 \times$ rms). The noise
contribution to this spread is negligible according to our synthetic spectra
model ($3 \times rms = 0.013$ km s$^{-1}$), so we can consider that
the scatter represents true velocity variations. It is possible
that part of this dispersion arises from a global component 
(e.g., rotation), but the results in Fig. 12 show that this
component cannot be dominant. Crudely assuming 
equipartition, we estimate the internal 
core motions to be at least 0.05 km s$^{-1}$, although they can be as
high as 0.1 km s$^{-1}$.

\begin{figure}
\centering
\resizebox{\hsize}{!}{\includegraphics{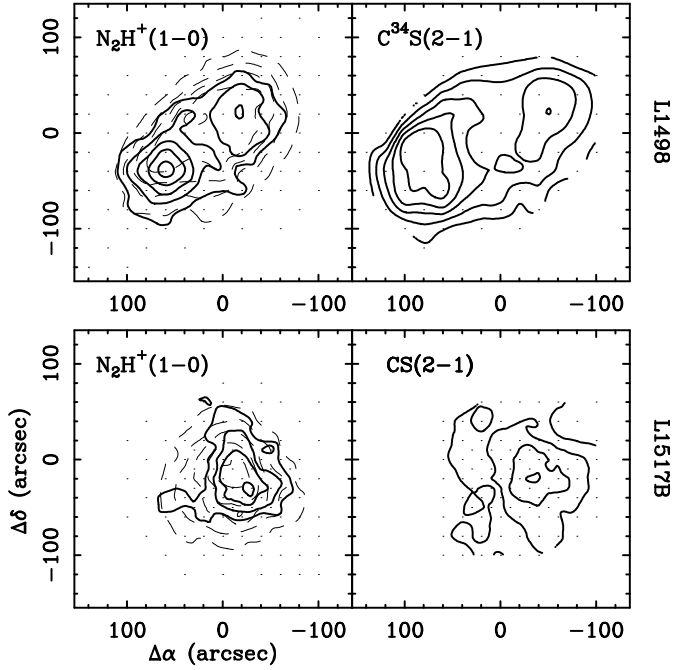}}
\caption{
Comparison of ``high velocity'' N$_2$H$^+$ emission to C$^{34}$S or 
CS emission showing their similar distribution. This is an
indication that the gas velocity structure is
related to the deviations from spherical symmetry 
in the pattern of molecular freeze out, probably due
to asymmetries in the core contraction history (see text). The solid lines in 
the left panels are N$_2$H$^+$ integrated intensity maps
for velocity intervals 0.2 km s$^{-1}$ (one linewidth, see Fig. 10)
away from the line center velocity (towards blue for L1517B and towards 
red for L1498) and 0.2 km s$^{-1}$ wide. The C$^{34}$S or CS maps 
in the right panels and
the full integrated N$_2$H$^+$ map (dashed) are as in Fig. 3.
\label{fig13}}
\end{figure}

Although the gas motions are small and subsonic (sound speed is 0.19 km 
s$^{-1}$ for 10 K), and therefore have little effect on the 
core energy budget, 
their presence indicates that the cores have not yet attained  
a state of perfect hydrostatic equilibrium. How these motions originate 
is unclear, but several indications suggest that they are
related to the process of core formation. First, the motions
are similar in magnitude to the inward motions 
observed in other cores \citep{lee99b,lee01}, and these are believed 
to be associated with
core contraction (note that L1498 has also infall asymmetry, while
L1517B has the opposite). Second, the 
time scale of these motions is of the order
of 1 Myr (0.05 km s$^{-1}$ over 0.05 pc or $75''$), a typical
starless core lifetime \citep{lee99a}. Finally, and more important, 
there is a correlation between
some velocity components and the non azimuthal asymmetries in the 
abundance of the depleted molecules (CS and CO). These abundance asymmetries
are most likely related to different contraction times of different
parts of the core (section 4.3), and thus their correlation with the velocity
pattern suggests that the pattern is related to contraction.
To show this, we present in Figure 13 (left panels)
the distribution of ``high velocity'' N$_2$H$^+$ emission (blue for L1498
and red for L1517B) superposed upon the integrated N$_2$H$^+$ 
emission (see caption for details). As the right panels in 
the figure show, these ``high velocity'' N$_2$H$^+$ components have 
distributions remarkably close to the distribution of bright 
CS/C$^{34}$S spots (more so than to CO probably because CS and
N$_2$H$^+$ are sensitive to similar densities), which we have seen most 
likely arise from differences in the amount of CS freeze-out. 
The similar distributions
are more remarkable because CS and CO hardly coexist 
with N$_2$H$^+$, as they deplete at radii for which N$_2$H$^+$
is non-detectable, and suggest a correlation of high CS
abundance with the ``high velocity'' components.
As we will see below, these high CS-abundance spots most likely arise 
from material that has been recently acquired by
the core and has therefore been at high density
for shorter time than the rest (i.e., it is ``younger'' and less depleted, 
see also \citealt{kui96}).
A peculiar velocity in this material is exactly what would have been 
expected if this interpretation is correct.

\section{Discussion}

\subsection{Core structure and equilibrium state}

We have seen that the systematic internal motions in both L1498 and 
L1517B are of the order of 0.1 km s$^{-1}$ and therefore subsonic
for gas at 10 K (sound speed is 0.19  km s$^{-1}$).
This suggests that both cores are close to a state of hydrostatic 
equilibrium in which the core self gravity is balanced by internal
forces. Given the core parameters we have derived in previous
sections, we now analyze whether such an equilibrium state is possible
and whether it is consistent with the observed core structure.

The simplest support component in each core is the thermal pressure
$P_T = n k T$, where $n$ is the gas density, 
$k$ is the Boltzmann constant, and $T$ the gas kinetic temperature.
This pressure may be supplemented by a
non-thermal component $P_{NT} = m n \sigma_{NT}^2$
if the non-thermal linewidth arises from turbulent supporting motions
($m$ is the mean particle mass and $\sigma_{NT}$ is the non thermal velocity 
dispersion, related to the non thermal FWHM by 
$\Delta V_{NT} = \sqrt{8\; ln 2}\; \sigma_{NT}$). The relative
importance of these two supporting components is given by the
ratio $P_{NT}/P_T = \sigma_{NT}^2/(k T/m)$, which can
be easily estimated from the core parameters derived in previous sections.
As both $\sigma_{NT}$ and $T$ are constant over the central
0.1 pc of L1498 and L1517B, the pressure ratio is constant, and has a value of
0.07 for $\Delta V_{NT} = 0.12$ km s$^{-1}$ and $T=10$ K (Sections 3.3 and 
3.7.1). This low ratio indicates that the non thermal motions contribute 
negligibly to the core support over at least the central 0.1 pc 
(see also \citealt{ful93}), and that in the absence of
support by an ordered magnetic field, thermal pressure is 
the main supporting agent in both L1498 and L1517B
(a similar conclusion for the case of B68 has been recently
obtained by \citealt{lad03}).

If thermal pressure dominates core support and the gas temperature is 
constant, the equation of core hydrostatic equilibrium reduces to the 
isothermal case, which has a classical solution for spherical geometry
\citep{cha39}. The addition of a constant non-thermal contribution 
implies only a simple generalization of the isothermal case and
does not change the solution significantly (e.g., 
Appendix A of \citealt{lau96}). As we have seen in section 3.1, 
the density profiles derived from the dust continuum emission do in fact
have {\em shapes} very close to those expected for isothermal 
spheres, especially in the case of L1517B. To check whether 
these profiles are consistent with equilibrium between self gravity 
and internal (thermal plus non thermal) pressure, we can compare the measured
velocity dispersion $\sigma$ (=0.185 km s$^{-1}$ for gas at  9.5 K
and $\Delta V_{NT} = 0.11$ km s$^{-1}$) with the velocity dispersion
required by the isothermal fit to the dust emission (section 3.1).
Unfortunately, the fit only constrains the combined parameter
$\sigma \sqrt{\kappa_\nu \; B(T_d)}$, making impossible a full check
without independent estimates of the dust emissivity $\kappa_\nu$ and 
temperature $T_d$. 
\footnote{The parameters directly derived from 
fitting the radial distribution of
dust emission with an isothermal model (including an on-the-fly simulation)
are the King radius $r_K = \sqrt{9\sigma^2/4\pi G \rho_0}$ (from the shape
of the profile) and the central emissivity $j_\nu = \kappa_\nu \rho_0 B(T_d)$
(from the intensity scale),
where $\rho_0$ is the central density and $T_d$ the dust temperature.
Substituting $\rho_0$ from the second equation into the first one,
we obtain $\sigma \sqrt{\kappa_\nu \; B(T_d)}$.} 
To take into account this dependence, we
re-write the velocity dispersions required from the continuum fits 
in Table 1 (where we assumed $\kappa_\nu=0.005$ cm$^2$ g$^{-1}$ 
and $T_d=10$ K) as
$\sigma = 0.27\; (\kappa_\nu/0.005)^{-0.5} \; (B(T_d)/B(10))^{-0.5}$ 
km s$^{-1}$
for L1517B and $0.32\; (\kappa_\nu/0.005)^{-0.5} \; (B(T_d)/B(10))^{-0.5}$ 
km s$^{-1}$ for L1498.  These values show that if 
$\kappa_\nu = 0.005$ cm$^2$ g$^{-1}$ and $T_d=10$ K, 
there is either an extra pressure contribution
which does not affect the observed linewidth or the cores should be
contracting under their own gravity. 

Although a state
of slow contraction is likely in both L1498 and L1517B (section 3.7.2),
we cannot rule out that at least part of 
the velocity dispersion mismatch reflects a wrong choice
of dust parameters. For example, the value of 
$T_d=10 K$, chosen so it equals the well-measured gas temperature,
may in fact be a slight overestimate. This is suggested by the 
models of \citet{gal02} (their Fig.3), which show that although 
dust and gas are thermally coupled at core densities the dust is
cooler by about 2 K. A lower dust temperature requires a larger
$\sigma$, and this makes the pressure unbalance more severe, although only by a 
small margin: a dust temperature of 8 K requires a 1.2 increase in
$\sigma$. More critical in the balance argument is the choice of
the dust emissivity $\kappa_\nu$, as this parameter is considered to 
be uncertain by a factor of 2. In L1517B, for example, 
if $\kappa_\nu = 0.01$ cm$^2$ g$^{-1}$, the gas will be in hydrostatic 
equilibrium assuming $T_d=10 K$. This higher $\kappa_\nu$ value
is not too far from the 0.008 cm$^2$ g$^{-1}$ recommended  by
\citet{oss94} for a standard gas-to-dust ratio of 100 
and dust in regions of density $\approx 10^5$ cm$^{-3}$, which is
typical of the L1517B center, and recent observations do in fact suggest
a systematic increase of $\kappa_\nu$ at high densities 
\citep{kra03,bia03}.
Thus, it is still
possible that L1517B is exactly or very close to 
a state of hydrostatic equilibrium
where its (mostly thermal) pressure balances its self gravity.
This higher $\kappa_\nu$ value, however, will not bring L1498 to 
a state of equilibrium, and this is also consistent with the non spherical 
shape of this core. 

\subsection{Core chemistry and comparison with models}

The abundance profiles of NH$_3$, N$_2$H$^+$, and
the isotopomers of CO and CS can be qualitatively understood
as the result of the selective freeze out of molecules onto
cold dust grains as the gas becomes centrally concentrated during
core formation. The lower binding energy of N$_2$ to grain surfaces
makes this species relatively overabundant at high densities,
and this in turn favors the presence of molecules composed of N and H
(like NH$_3$ and N$_2$H$^+$) in the core interiors. This selective freeze
out of molecules during core contraction and its resulting chemistry
has been modeled in detail by a number of authors over the last few years
\citep{ber97,cha97,cas99,nej99,aik01,li02,she03,aik03}, and their results
are in general agreement with the observations and analysis
presented here. A detailed comparison between models and data, however,
reveals some differences between observed molecular behavior and theoretical
expectations, and in this section we comment on the possible origin of these
differences.

The main disagreement between our observations and the published models
involves the  behavior of NH$_3$ and N$_2$H$^+$ near the
core centers. As seen in sections 3.3 and 3.4, both L1498 and L1517B
present the same pattern of constant N$_2$H$^+$ abundance
together with a factor-of-a-few central NH$_3$ increase, and this
pattern seems also present in other starless cores (see TMCWC).
Most chemical models, however, predict that NH$_3$ and N$_2$H$^+$ will behave
in a similar fashion given that they have N$_2$ as a common parent.
This is qualitatively correct in that the two species trace
the same regions but   the quantitative differences noted above
require explanation. \citet{aik03} have made a detailed
attempt to fit observations of L1544 (their Fig. 8). Their best
model (a delayed collapse with N initially molecular and a high
CO binding energy) has an essentially constant NH$_3$/N$_2$H$^+$ abundance
ratio as a function of radius.

It is unclear which processes cause
 the NH$_3$ and N$_2$H$^+$ abundances to follow
different trends at high densities
and produce the relative enhancement of NH$_3$. Both have
  their origins in molecular nitrogen and hence their abundance
ratio should be independent of the N$_{2}$ abundance. Their
formation involves reactions with He$^{+}$ (in the case of
NH$_{3}$) and H$^{+}_{3}$ (in the case of N$_2$H$^+$ ) and
thus their abundance ratio is sensitive to the abundance ratio
of these ions. The ionic abundances
 are certainly sensitive to the depletion
of CO and other molecules  and hence uncertainties in key rates determining
the abundances of He$^{+}$ and H$_3^+$ may be responsible for
variations in the NH$_{3}$/N$_{2}$H$^+$ abundance ratio. Apart from this,
 we note that while N$_{2}$H$^{+}$ is likely mainly destroyed
by recombination with electrons and reactions with CO,
it is rather unclear what the main destruction mechanism for
NH$_{3}$ is in regions where CO (and hence carbon compounds in
general) is sufficiently depleted that reactions with ions
such as C$^{+}$ become negligible. The answers to these questions
may provide the explanation for the observed gradient in
the NH$_{3}$/N$_2$H$^+$ ratio.

\subsection{Core formation}

Although it is clear that dense
cores contract from the lower density gas of their
surrounding molecular clouds, how this process occurs and
how long it takes is still a matter of much uncertainty (see, e.g., 
\citealt{shu87} and \citealt{har01} for two opposing views).
This state of affairs results in part because it is not yet possible
to infer a core contraction history directly from observations,
as we lack reliable time markers to assess the evolutionary state
of different cores. Fortunately, the time and density dependence of the 
freeze out process may offer the possibility of shedding some
light on the core contraction history by allowing us to 
correlate the molecular abundances of volatile species with the 
evolutionary state of the gas in a core. Here we present a first
(and preliminary)
application of these ideas to the case of the L1498 and L1517B cores.

Both L1498 and L1517B appear highly symmetric in their distribution
of N$_2$H$^+$, NH$_3$, and dust continuum (Fig. 3), which are
the most reliable tracers of the dense core gas. L1498,
especially in the density-sensitive N$_2$H$^+$, presents an evident
bilateral symmetry with respect to a NE-SW axis, and L1517B has
an almost perfect circular symmetry with only minor deviations.
These symmetries in the emission of the most robust tracers most likely 
reflect a high degree of symmetry in the underlying distributions of
matter in the cores, in agreement with their state close to hydrostatic
equilibrium.

In contrast with the symmetric distributions of
N$_2$H$^+$, NH$_3$, and dust continuum, the distributions
of CS and CO emission in L1498 and L1517B are highly asymmetric (Fig. 3).
As we have seen in sections 3.5 and 3.6, 
these asymmetries arise from a pattern of 
non axisymmetric  abundance variations that is
superposed to 
the central abundance drop due to molecular freeze out.
These non axisymmetric variations are
located at radii where freeze out is already 
prevalent, so they very likely represent 
variations in the amount of freeze out, in the sense that
the bright spots to the SE and E of L1498 and to the W of L1517B
are places where this process is less severe and therefore 
the CO and CS abundances are  higher.
This interpretation is supported by the fact that the same regions have
at the same time
bright CO, CS, and other molecules that freeze out at similar densities 
(Tafalla et al. 
2004 in preparation), indicating that the process responsible for the
asymmetry is not
limited to one particular chemical species. 

As molecular freeze out is very sensitive to 
the time the gas has spent at high density (e.g., \citealt{ber97}), the
simplest explanation of the lower freeze out at the CO and CS
bright spots is that these regions
have somehow stayed at high density
less time than other regions with similar radius
(and therefore similar present density). (Pre-existing chemical 
inhomogeneities seem unlikely because of the small scale.)
For this to happen, the contraction of the gas in L1498 and 
L1517B has to have been non spherically symmetric, with the gas
at the CO and CS bright spots being ``younger''
(less processed) and therefore more recently accreted.
This interpretation is consistent with the correlation between
enhanced CO/CS abundance and ``high'' velocity N$_2$H$^+$ and NH$_3$ 
found in
section 3.7.2 (see Fig. 13), as the younger portions of the core will 
have not yet attained
perfect equilibrium and thus retain a fraction of their contraction speed.
Kinematics and chemical composition, therefore, agree in suggesting that
L1498 and L1517B have formed from non spherical contraction
of cloud material. The time scale of this contraction seems to be
of the order of a few Myr (section 3.7.2).

The above time scale, although short, is not enough to discriminate
between the different scenarios of star formation. On
the one hand, the value is
consistent with a picture of relatively rapid star formation 
\citep{elm00,har01}, but on the other, 
it is also consistent with ambipolar diffusion models
that start with an initially not highly subcritical
cloud \citep{li99,cio01}.
In paper 2, we will revisit the time scale issue based on
the analysis of the freeze out process of a large number of molecular 
species. Here we note that if the magnetically-mediated scenario
is correct,
the fastest gas motions are expected to occur along the 
magnetic field axis, and this would imply  a correlation of the 
observed abundance anisotropies with the magnetic field direction.
Observations of the polarization direction in the submillimeter continuum
should be able to test this prediction.

More conclusive than the kinematic estimate of the contraction
time scale is the measurement of the (low) turbulence level
inside the cores. Our observations 
of L1498 and L1517B (Fig. 10, top) appear to be inconsistent with 
typical expectations from the supersonic 
turbulence scenario (see \citealt{mac03} for a recent review). 
\citet{bal03} have recently argued that density profiles similar to
those of hydrostatic equilibrium systems can appear fortuitously in 
model simulations of supersonic turbulence, showing that the shape of the
density profile is not a reliable indication of a core equilibrium state. 
Although it is true that the cores in the \citet{bal03} simulations
can mimic the density profiles of systems like L1498 and L1517B,
their internal velocity fields clearly betray their ``hydrostatic 
disguise.'' Turbulent model cores in the simulations of 
\citet{bal03} present velocity profiles with internal changes larger than 
the speed of sound, and on the order of 0.5-1 km s$^{-1}$
in cases of driven turbulence (their Figs. 9-11). Our observations
rule out such strong velocity changes by about an order of 
magnitude, and not only in the plane of
the sky (from our radial profiles of line center velocity, see
Fig. 11), but also along any given line of sight (our linewidth measurements,
see Fig. 10).

The problem of the supersonic turbulence scenario 
is unlikely to be restricted to L1498 and L1517B, as these objects 
seem representative of the
Taurus core population in, for example, having narrow lines 
(e.g., \citealt{ful93}). This problem points to the need of any model
of core formation for producing condensations which appear
thermally dominated and extremely quiescent. How these
close to hydrostatic equilibrium configurations can be achieved
in the relatively short time of a few Myr seems to 
pose a challenge to most core formation models.

\section{Conclusions}

We have presented detailed models of the internal structure of two 
close-to-round starless cores in the nearby Taurus-Auriga star formation
complex, L1498 and L1517B. Our models have been constrained from the
simultaneous fit of the spatial distribution and the central emerging
spectrum of at least two transitions of NH$_3$, N$_2$H$^+$, CS,
C$^{34}$S, C$^{18}$O, and C$^{17}$O, together with maps of the 1.2 mm
continuum.  The main conclusions of this work are:

1. Both cores present gas temperature distributions consistent with 
a constant value (10 K for L1498 and 9.5 K for L1517B) and a very
small rms (1 K). No evidence for radial temperature gradients
is found in the central 0.1 pc.

2. Assuming constant dust temperature and emissivity, we derive density
profiles for both L1498 and L1517B of the form
$n(r) = n_0/(1+(r/r_0)^\alpha)$, where $n_0$, $r_0$, and $\alpha$ 
are free parameters. For L1517B, the $\alpha=2.5$ best fit is 
indistinguishable from an isothermal sphere fit, and for L1498
the $\alpha=3.5$ best fit is close to the isothermal fit.

3. From the combination of radiative transfer modeling and hyperfine
fitting of the NH$_3$ and N$_2$H$^+$ spectra, we determine that in
both L1498 and L1517B the
non thermal component of the linewidth is constant with radius
over at least the central 0.1 pc. This component
is smaller than the thermal linewidth and has a FWHM of about 0.12 km 
s$^{-1}$ in both cores. Its scatter in the radial profiles
is extremely small ($<5$\%), and it is consistent with 
the measurement uncertainty.

4. Using a Monte Carlo radiative transfer code and 
the radial profiles of density, temperature, and linewidth,
we determine radial profiles of molecular abundance for all observed
species. For NH$_3$, the abundance at the core centers is a 
factor of a few higher than at intermediate radii (0.05 pc). For N$_2$H$^+$,
a constant abundance model fits well the emission at all radii
in L1517B, while a drop by
a factor of 3 at large radii is preferred in L1498. For both CO 
and CS (and isotopomers), constant abundance models fail dramatically
to fit the emission, and an abrupt central abundance drop is needed
to explain the data. The abundance drop of these two species is so severe
that the emission can be modeled with step functions and no
molecules at the center.

5. The central abundance drop of CO and CS
seems to arise from the freeze out of these molecules onto cold
dust grains at densities of a few $10^4$ cm$^{-3}$.  The simultaneous 
survival of 
N$_2$H$^+$ and NH$_3$ seems related to the 
lower binding energy to grains of N$_2$, which allows
the production of these two species 
at the core centers. The relative increase of the NH$_3$ abundance
over N$_2$H$^+$ probably results from a combination of a more 
efficient NH$_3$ production in the highly CO-depleted core interiors
and a decrease in the ion fraction with density, although further
modeling is required to understand this behavior.

6. Superposed to the pattern of radial abundance gradients
there is a weaker pattern of non axially symmetric abundance 
variations in both CO and CS. 
This pattern results from asymmetries in the
distribution of CO and CS freeze out, and its
presence indicates that
the process responsible for the freeze out has not been 
spherically symmetric. As gas contraction to form the cores
is the likely cause of the central freeze out, the asymmetry suggests
that core formation has occurred in a non spherically symmetric way.

7. The analysis of the line center velocities of the freeze-out resistant
N$_2$H$^+$ and NH$_3$
reveals spatial variations at the level of 0.1 km s$^{-1}$
in both L1498 and L1517B. The spatial distribution of these variations is
complex and cannot be explained with a simple pattern
like rotation. It most likely represents internal bulk motions
in the core gas. Some gas at anomalous velocities has a spatial 
distribution similar to the gas with anomalously high
CO and CS abundance in the pattern of non axisymmetric freeze out.
This similarity suggests that part of the observed bulk motions 
result from residual core contraction. If so, the contraction time is 
of the order of 1 Myr.

8. We have used the model physical parameters to study the stability of 
L1498 and L1517B.  The small non thermal linewidth 
indicates that any turbulent pressure support is negligible in both cores
($<10$ \%), and that in the absence of magnetic fields, 
thermal pressure is the only stabilizing agent. Although the density profiles
have the shape expected for isothermal equilibrium, the mass estimated from
the continuum is larger than what could be held in equilibrium. Either
our choice of dust parameters is incorrect, there is additional magnetic
support, or the cores are in state of (slow) contraction.
In any case, the observed velocity and linewidth profiles
in both L1498 and L1517B are inconsistent with representative 
predictions 
of supersonic turbulence models
that generate density profiles similar to those of an isothermal sphere.

\acknowledgements
We thank the IRAM 30m staff for help during the observations, 
Carl Gottlieb for communicating new frequency measurements
prior to publication,
Daniele Galli for providing us with model gas temperature 
profiles for L1517B, 
Zhi-Yun Li for advise on ambipolar diffusion models,
and Alexander Lapinov for bringing to our attention
recent work on approximations to the isothermal sphere.
We also thank the referee, Ted Bergin, for comments that 
helped clarifying the manuscript.
MT acknowledges partial support from grant AYA2000-0927 of
the Spanish DGES. PC and CMW acknowledge support from the MIUR project ``Dust
and Molecules in Astrophysical Environments."
This research has made use of NASA's Astrophysics Data System
Bibliographic Services and the SIMBAD database, operated at CDS,
Strasbourg, France.

\appendix

\section{A simple analytic approximation to the isothermal sphere}

In TMCWC we fitted the volume density profiles of 5 dense cores using an 
analytical expression of the form
$$n(r) = {n_0\over 1+(r/r_0)^\alpha}, $$ 
with $n_0$, $r_0$, and $\alpha$ as free parameters. The motivation of this
choice was purely empirical, based on the ability of these profiles
to fit the observed radial distribution of dust continuum emission.
A later hydrostatic equilibrium analysis of these profiles (unpublished)
showed that for $\alpha=2.5$ an almost perfect balance between isothermal 
plus constant turbulence pressure and self gravity was achieved. Intrigued 
by this ``coincidence,'' especially because an $\alpha=2.5$ profile lacks
the proper asymptotic behavior for an isothermal sphere $(1/r^2)$, we 
carried out a point-by-point
comparison between this profile and the numerical solution of the isothermal
function given by \citet{cha49}. These authors calculated
numerically the mass density profile for an isothermal sphere normalized to 
the central density ($\rho_0$), 
which can be written as $\rho=\exp(-\psi)$ with $\psi$ being the solution of
$${1\over \xi^2} {d\over d \xi}\left(\xi^2 {d\psi \over d\xi}\right) = 
e^{-\psi},$$
and $\xi = \sqrt{4\pi G \rho_0/\sigma^2}\; r$ (see \citealt{cha39} for a 
full discussion). To compare our $\alpha=2.5$ profile with this function, we 
normalize it and require that it reaches the half maximum density at the same
radius as the isothermal solution ($\xi \approx 2.25$), and this 
completely fixes the profile as 
$$\rho_{app}(\xi) = {1\over {1+(\xi/2.25)^{2.5}}}$$.

\begin{figure}
\centering
\resizebox{\hsize}{!}{\includegraphics{4112.fa1}}
\caption{Top: Numerical solution of the isothermal function from 
\citet{cha49} (squares) compared to our simple $\alpha=2.5$
analytic model (line). Bottom: relative error of several
analytic approximations to the isothermal sphere. The solid line 
represents the $\alpha=2.5$ profile presented here, the long-dashed lines
is the modified Hubble law, and the dotted line is model b from
\citet{nat97}. Note how the $\alpha=2.5$ profile remains
within 10\% of the exact solution to higher $\xi$ values than the other 
profiles (although it finally diverges while the \citet{nat97}
curve converges to the right solution at large $\xi$).
\label{figa1}}
\end{figure}

	Fig. A1 shows (top) a comparison between the $\alpha=2.5$ profile
and the numerical solution from \citet{cha49}.
For additional comparison, we also present the classical modified Hubble law
approximation \citep{bin87}
and the recent simple approximation
proposed by \citet{nat97} (their case b). As can be seen,
the $\alpha=2.5$ profile remains within 10\% from the exact solution
up to a larger $\xi$ value than the other solutions (although 
the \citet{nat97} profile finally converges to the
exact solution for large $\xi$). In fact, the $\alpha=2.5$ solution
is accurate within 10\% for $\xi \le 23$,
more than 3.5 times the instability radius \citep{ebe55,bon56}, 
and over this range, the density varies by more than a factor 300.
The associated instability point of this solution ($\xi=6.467$) is 
within 0.25\% of the exact solution ($\xi=6.451$, \citealt{hun77}).
This shows that the $\alpha=2.5$ profile can be used to approximate 
any stable Bonnor-Ebert sphere within 10\%.

\section{Kinetic temperature estimate from NH$_3$}

The lack of allowed radiative transitions between the (J,K)=(2,2) and 
(1,1) levels
of NH$_3$ makes their relative population (described by the rotational 
temperature $T_R^{21}$) highly sensitive to collisions, and therefore a 
good estimator of the gas kinetic temperature. In the traditional NH$_3$
analysis, the (1,1) and (2,2) populations are calculated from the 
spectra assuming LTE
conditions, and the derived $T_R^{21}$ rotational temperature is converted
into an estimate of the gas kinetic temperature $T_K$ \citep{wal83}.
To test this procedure and to calibrate the 
$T_R^{21}$-$T_K$ relation using realistic radiative transfer and 
cloud conditions, we have run a series of Monte Carlo models
and produced synthetic (1,1) and (2,2) spectra. These spectra 
have been analyzed using the standard NH$_3$ procedure, and the derived
gas parameters have been compared with those used in the models.

Our NH$_3$ Monte Carlo model has been described in TMCWC and is based on that
of \citet{ber79}, adapted to handle NH$_3$. It includes the 6 lowest 
levels of para-NH$_3$, which contain more than 99.9\% of the molecules at 
10 K (typical core gas temperature). It assumes that the relative hf 
sub-populations of the meta-stable
levels are in LTE (as required by observations), and calculates the 
relative populations between the (1,1), (2,1), and (2,2) levels 
in a self-consistent manner. It uses 
the collision coefficients with H$_2$ of \citet{dan88}, which seems to
be the most accurate set available, according to recent laboratory
measurements \citep{wil02}. As cloud
parameters, we use the density profiles derived from dust emission for L1498 
({\em L1498-models}) and L1517B ({\em L1517B-models}), and assume a constant
gas temperature (given the results of section 3.3), which we have varied from
5 to 20 K. 

In all 9 models we have run we find that while both the (1,1) and (2,2)
excitation temperatures vary with radius by a factor of several
(as a result of the core density gradient), 
the $T_R^{21}$ rotational temperature stays constant over the 
core within 2\% or less. This makes 
the derivation of $T_R^{21}$ using the standard NH$_3$ analysis
(e.g., \citealt{bac87}) rather accurate, despite of violating 
one of its basic assumptions: that of a single excitation temperature
for both (1,1) and (2,2). In fact, our models show that in the range 
of kinetic temperatures we
have explored (5-20 K), $T_R^{21}$ estimates using the standard analysis
agree within 5\% with the real values, and that the  
agreement is at the 1\% level for temperatures around 10 K or lower.
Thus, $T_R^{21}$ can be easily and accurately 
determined from observations for realistic core conditions.

\begin{figure}
\centering
\resizebox{7cm}{!}{\includegraphics{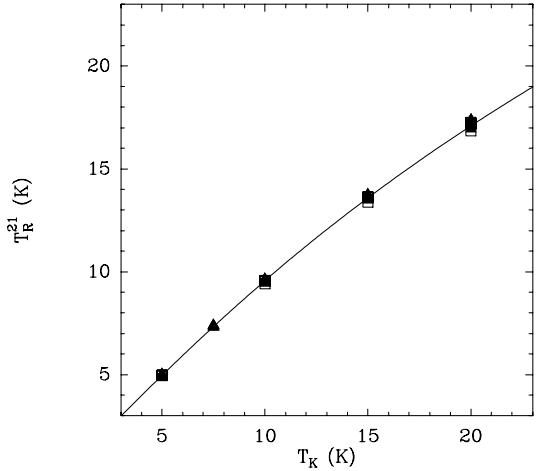}}
\caption{$T_R^{21}$ versus $T_K$ derived using Monte Carlo models.
$T_K$ is the (constant) gas kinetic temperature and $T_R^{21}$
is the (2,2)-to-(1,1) rotation temperature derived by applying the
standard NH$_3$ analysis to the Monte Carlo spectra. Filled triangles 
and open squares represent
models having the analytic density profile derived in section
3.1 for L1517B and L1498, respectively (see Table 2). Multiple 
points at the same $T_K$ represent values measured at radii equal
0, 10, 30, 60, 100, and 150 arcseconds (data convolved to the $40''$
resolution of the Effelsberg 100m telescope). The solid line is the
analytic expression proposed in the text to derive  $T_K$ using 
the $T_R^{21}$ estimate from the standard NH$_3$ analysis.
\label{figb1}}
\end{figure}

The critical part of the NH$_3$ analysis is the conversion of the $T_R^{21}$
estimate into an estimate of the gas kinetic temperature $T_K$, as
this step requires a full solution of the radiative transfer
equation. \citet{wal83} have presented a simple analytic 
expression derived under simplifying assumptions and using an old 
version of the NH$_3$-H$_2$ collision coefficients from \citet{gre81}. 
In order to derive a revised expression using realistic radiative
transport and the new collision coefficients of \citet{dan88},
we compare the $T_K$ inputs in the Monte Carlo models
with the $T_R^{21}$ values derived from the above analysis.
As can be seen in Figure B.1., for any given $T_K$ between 5 and 20K, the
L1498-models and L1517B-models predict the same $T_R^{21}$ within
5\% or better, and this shows that at least for our conditions, the 
$T_R^{21}$-$T_K$ relation is almost independent of the
details of core density, size, etc.
To derive an analytic expression for this relation, we use as
inspiration equation 2 of \citet{wal83}, although we look
for the inverse function of theirs
because we want to predict $T_K$ given the $T_R^{21}$
value determined with the standard NH$_3$ analysis. After some 
experimentation, we have found that the following expression
$$T_K = {T_R^{21}\over 1-{T_R^{21}\over 42}\; \ln[1+1.1\exp(-16/T_R^{21})]}$$
fits the points in the range $T_K=$ 5-20 K to better than 5\%, and therefore
provides an accurate gas temperature estimate based on 
easy-to-measure quantities. We recommend this relation to derive
$T_K$ in dense, cold cores at temperatures lower than or around 20~K 
(extrapolation to higher temperatures requires further modeling).

\bibliographystyle{apj}

\end{document}